\PassOptionsToPackage{unicode}{hyperref}
\PassOptionsToPackage{hyphens}{url}
\documentclass[
]{article}
\usepackage{amsmath,amssymb}
\usepackage{iftex}
\ifPDFTeX
  \usepackage[T1]{fontenc}
  \usepackage[utf8]{inputenc}
  \usepackage{textcomp} 
\else 
  \usepackage{unicode-math} 
  \defaultfontfeatures{Scale=MatchLowercase}
  \defaultfontfeatures[\rmfamily]{Ligatures=TeX,Scale=1}
\fi
\usepackage{lmodern}
\ifPDFTeX\else
\fi
\IfFileExists{upquote.sty}{\usepackage{upquote}}{}
\IfFileExists{microtype.sty}{
  \usepackage[]{microtype}
  \UseMicrotypeSet[protrusion]{basicmath} 
}{}
\makeatletter
\@ifundefined{KOMAClassName}{
  \IfFileExists{parskip.sty}{%
    \usepackage{parskip}
  }{
    \setlength{\parindent}{0pt}
    \setlength{\parskip}{6pt plus 2pt minus 1pt}}
}{
  \KOMAoptions{parskip=half}}
\makeatother
\usepackage{xcolor}
\usepackage[margin=1in]{geometry}
\usepackage{color}
\usepackage{fancyvrb}

\DefineVerbatimEnvironment{Highlighting}{Verbatim}{commandchars=\\\{\}}
\usepackage{framed}
\definecolor{shadecolor}{RGB}{248,248,248}
\newenvironment{Shaded}{\begin{snugshade}}{\end{snugshade}}

\newcommand{\AttributeTok}[1]{\textcolor[rgb]{0.13,0.29,0.53}{#1}}

\newcommand{\CommentTok}[1]{\textcolor[rgb]{0.56,0.35,0.01}{\textit{#1}}}

\newcommand{\ConstantTok}[1]{\textcolor[rgb]{0.56,0.35,0.01}{#1}}

\newcommand{\DecValTok}[1]{\textcolor[rgb]{0.00,0.00,0.81}{#1}}
\newcommand{\DocumentationTok}[1]{\textcolor[rgb]{0.56,0.35,0.01}{\textbf{\textit{#1}}}}

\newcommand{\FloatTok}[1]{\textcolor[rgb]{0.00,0.00,0.81}{#1}}
\newcommand{\FunctionTok}[1]{\textcolor[rgb]{0.13,0.29,0.53}{\textbf{#1}}}

\newcommand{\NormalTok}[1]{#1}

\newcommand{\OtherTok}[1]{\textcolor[rgb]{0.56,0.35,0.01}{#1}}

\newcommand{\SpecialCharTok}[1]{\textcolor[rgb]{0.81,0.36,0.00}{\textbf{#1}}}

\newcommand{\StringTok}[1]{\textcolor[rgb]{0.31,0.60,0.02}{#1}}

\usepackage{graphicx}
\makeatletter
\def\maxwidth{\ifdim\Gin@nat@width>\linewidth\linewidth\else\Gin@nat@width\fi}
\def\maxheight{\ifdim\Gin@nat@height>\textheight\textheight\else\Gin@nat@height\fi}
\makeatother
\setkeys{Gin}{width=\maxwidth,height=\maxheight,keepaspectratio}
\makeatletter
\def\fps@figure{htbp}
\makeatother
\providecommand{\tightlist}{%
  \setlength{\itemsep}{0pt}\setlength{\parskip}{0pt}}
\ifLuaTeX
  \usepackage{selnolig}  
\fi
\IfFileExists{bookmark.sty}{\usepackage{bookmark}}{\usepackage{hyperref}}
\IfFileExists{xurl.sty}{\usepackage{xurl}}{} 
\hypersetup{
  pdftitle={Gaussian spatial regression using the spmoran package: case study examples},
  pdfauthor={Daisuke Murakami},
  hidelinks,
  pdfcreator={LaTeX via pandoc}}

\title{Gaussian spatial regression using the spmoran package: case study
examples}
\author{Daisuke Murakami}
\date{2023/11/01}

\begin{document}
\maketitle

{
\setcounter{tocdepth}{4}
\tableofcontents
}
\hypertarget{introduction}{%
\section{Introduction}\label{introduction}}

This package provides functions for estimating Gaussian and non-Gaussian
spatial regression models and extensions, including spatially and
non-spatially varying coefficient models, models with group effects,
spatial unconditional quantile regression models, and low rank spatial
econometric models. All these models are estimated computationally
efficiently.

An approximate Gaussian process (GP or kriging model), which is
interpretable in terms of the Moran coefficient (MC), is used for
modeling the spatial process. The approximate GP is defined by a linear
combination of the Moran eigenvectors (MEs) corresponding to positive
eigenvalue, which are known to explain positive spatial dependence. The
resulting spatial process describes positively dependent map patterns
(i.e., MC \textgreater{} 0), which are dominant in regional science
(Griffith, 2003). Below, the spmoran package is used to analyse the
Boston housing dataset.

\textcolor{blue}{The sample codes used below are available from <https://github.com/dmuraka/spmoran>}.
While this vignette mainly focuses on Gaussian regression modeling,
another vignette focusing on non-Gaussian regression modeling and count
regression modeling is also available from the same GitHub page (and
Murakami 2021).

\begin{Shaded}
\begin{Highlighting}[]
\FunctionTok{library}\NormalTok{(spmoran)}
\end{Highlighting}
\end{Shaded}

\hypertarget{gaussian-spatial-regression-models}{%
\section{Gaussian spatial regression
models}\label{gaussian-spatial-regression-models}}

\hypertarget{basic-models}{%
\subsection{Basic models}\label{basic-models}}

This section considers the following model:
\[y_i=\sum^K_{k=1}x_{i,k}\beta_k+f_{MC}(s_i)+\epsilon_i, \hspace{0.5cm}
\epsilon_i \sim N(0, \sigma^2),\] which decomposes the explained
variable \(y_i\) observed at the i-th sample site into trend
\(\sum^K_{k=1}x_{i,k}\beta_{i,k}\), spatial process \(f_{MC}(s_i)\)
depending on location \(s_i\), and noise \(\epsilon_i\). The spatial
process is required to eliminate residual spatial dependence and
estimate/infer regression coefficients \(\beta_k\) appropriately. ESF
and RE-ESF define \(f_{MC}(s_i)\) using the MC-based spatial process to
efficiently eliminate residual spatial dependence. These processes are
defined by the weighted sum of the Moran eigenvectors (MEs), which are
spatial basis functions (distinct map pattern variables; see Griffith,
2003).

\hypertarget{eigenvector-spatial-filtering-esf}{%
\subsubsection{Eigenvector spatial filtering
(ESF)}\label{eigenvector-spatial-filtering-esf}}

ESF specifies \(f_{MC}(s_i)\) using an MC-based deterministic spatial
process (see Griffith, 2003). Below is a code estimating the linear ESF
model. In the code, the meigen function extracts the MEs, and the esf
function estimates the model.

\begin{Shaded}
\begin{Highlighting}[]
\FunctionTok{require}\NormalTok{(spdep)}
\FunctionTok{data}\NormalTok{(boston)}
\NormalTok{y       }\OtherTok{\textless{}{-}}\NormalTok{ boston.c[, }\StringTok{"CMEDV"}\NormalTok{ ]}
\NormalTok{x       }\OtherTok{\textless{}{-}}\NormalTok{ boston.c[,}\FunctionTok{c}\NormalTok{(}\StringTok{"CRIM"}\NormalTok{,}\StringTok{"ZN"}\NormalTok{,}\StringTok{"INDUS"}\NormalTok{, }\StringTok{"CHAS"}\NormalTok{, }\StringTok{"NOX"}\NormalTok{,}\StringTok{"RM"}\NormalTok{, }\StringTok{"AGE"}\NormalTok{)]}
\NormalTok{coords}\OtherTok{\textless{}{-}}\NormalTok{ boston.c[,}\FunctionTok{c}\NormalTok{(}\StringTok{"LON"}\NormalTok{,}\StringTok{"LAT"}\NormalTok{)]}

\DocumentationTok{\#\#\#\#\#\#\#\#\#Distance{-}based ESF}
\NormalTok{meig    }\OtherTok{\textless{}{-}} \FunctionTok{meigen}\NormalTok{(}\AttributeTok{coords=}\NormalTok{coords)}
\NormalTok{res   }\OtherTok{\textless{}{-}} \FunctionTok{esf}\NormalTok{(}\AttributeTok{y=}\NormalTok{y,}\AttributeTok{x=}\NormalTok{x,}\AttributeTok{meig=}\NormalTok{meig, }\AttributeTok{vif=}\DecValTok{10}\NormalTok{)}
\NormalTok{res}
\end{Highlighting}
\end{Shaded}

\begin{verbatim}
## Call:
## esf(y = y, x = x, vif = 10, meig = meig)
## 
## ----Coefficients------------------------------
##                 Estimate         SE    t_value      p_value
## (Intercept)  11.34040959 3.91692274  2.8952344 3.968277e-03
## CRIM         -0.20942091 0.03048530 -6.8695702 2.089395e-11
## ZN            0.02322000 0.01384823  1.6767492 9.426799e-02
## INDUS        -0.15063613 0.06823776 -2.2075188 2.776856e-02
## CHAS          0.15172838 0.93842988  0.1616832 8.716260e-01
## NOX         -38.02167637 4.79403898 -7.9310320 1.651338e-14
## RM            6.33316024 0.36887955 17.1686403 1.842211e-51
## AGE          -0.07820247 0.01564970 -4.9970593 8.274067e-07
## 
## ----Spatial effects (residuals)---------------
##                       Estimate
## SD                   6.8540461
## Moran.I/max(Moran.I) 0.6701035
## 
## ----Error statistics--------------------------
##                  stat
## resid_SE     4.476459
## adjR2        0.762328
## logLik   -1453.376154
## AIC       2996.752308
## BIC       3186.946458
\end{verbatim}

While the meigen function is slow for large samples, it can be
substituted with the meigen\_f function performing a fast
eigen-approximation. Here is a fast ESF code for large samples:

\begin{Shaded}
\begin{Highlighting}[]
\NormalTok{meig\_f}\OtherTok{\textless{}{-}} \FunctionTok{meigen\_f}\NormalTok{(coords)}
\NormalTok{res   }\OtherTok{\textless{}{-}} \FunctionTok{esf}\NormalTok{(}\AttributeTok{y=}\NormalTok{y, }\AttributeTok{x=}\NormalTok{x, }\AttributeTok{meig=}\NormalTok{meig\_f,}\AttributeTok{vif=}\DecValTok{10}\NormalTok{, }\AttributeTok{fn=}\StringTok{"all"}\NormalTok{)}
\end{Highlighting}
\end{Shaded}

\hypertarget{random-effects-esf-re-esf}{%
\subsubsection{Random effects ESF
(RE-ESF)}\label{random-effects-esf-re-esf}}

RE-ESF specifies \(f_{MC}(s_i)\) using an MC-based spatial random
process, again to eliminate residual spatial dependence (see Murakami
and Griffith, 2015). Here is a sample example:

\begin{Shaded}
\begin{Highlighting}[]
\NormalTok{res   }\OtherTok{\textless{}{-}} \FunctionTok{resf}\NormalTok{(}\AttributeTok{y =}\NormalTok{ y, }\AttributeTok{x =}\NormalTok{ x, }\AttributeTok{meig =}\NormalTok{ meig)}
\NormalTok{res}
\end{Highlighting}
\end{Shaded}

\begin{verbatim}
## Call:
## resf(y = y, x = x, meig = meig)
## 
## ----Coefficients------------------------------
##                 Estimate         SE    t_value      p_value
## (Intercept)   6.63220299 3.94484191  1.6812342 9.340110e-02
## CRIM         -0.19815203 0.03126666 -6.3374866 5.608678e-10
## ZN            0.01453736 0.01591772  0.9132814 3.615764e-01
## INDUS        -0.15560251 0.06842940 -2.2739131 2.343446e-02
## CHAS          0.51046253 0.92329946  0.5528678 5.806244e-01
## NOX         -31.26689956 5.02069120 -6.2276086 1.075126e-09
## RM            6.33993147 0.36671337 17.2885202 0.000000e+00
## AGE          -0.06351411 0.01526957 -4.1595218 3.810683e-05
## 
## ----Variance parameter------------------------
## 
## Spatial effects (residuals):
##                      (Intercept)
## random_SD              6.7424429
## Moran.I/max(Moran.I)   0.6648678
## 
## ----Error statistics--------------------------
##                      stat
## resid_SE        4.3515211
## adjR2(cond)     0.7735912
## rlogLik     -1540.3812428
## AIC          3102.7624855
## BIC          3149.2543889
## 
## NULL model: lm( y ~ x )
##    (r)loglik: -1612.825 ( AIC: 3243.65,  BIC: 3281.689 )
## 
## Note: AIC and BIC are based on the restricted/marginal likelihood.
##       Use method="ml" for comparison of models with different fixed effects (x)
\end{verbatim}

The residual spatial process \(f_{MC}(s_i)\) is plotted as follows:

\begin{Shaded}
\begin{Highlighting}[]
\FunctionTok{plot\_s}\NormalTok{(res)}
\end{Highlighting}
\end{Shaded}

\includegraphics{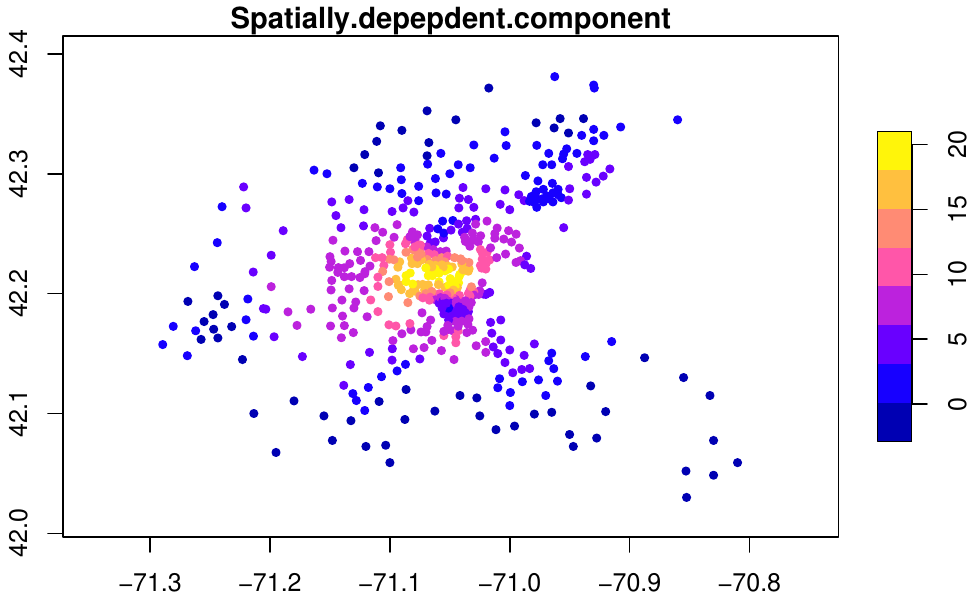}

For large data, the meigen\_f function is available again:

\begin{Shaded}
\begin{Highlighting}[]
\NormalTok{meig\_f}\OtherTok{\textless{}{-}} \FunctionTok{meigen\_f}\NormalTok{(coords)}
\NormalTok{res   }\OtherTok{\textless{}{-}} \FunctionTok{resf}\NormalTok{(}\AttributeTok{y =}\NormalTok{ y, }\AttributeTok{x =}\NormalTok{ x, }\AttributeTok{meig =}\NormalTok{ meig\_f)}
\end{Highlighting}
\end{Shaded}

The meigen\_f function is available for all the regression models
explained below.

\hypertarget{extended-models}{%
\subsection{Extended models}\label{extended-models}}

\hypertarget{models-with-non-spatially-varying-coefficients-coefficients-varying-wrt-covariate-value}{%
\subsubsection{Models with non-spatially varying coefficients
(coefficients varying wrt covariate
value)}\label{models-with-non-spatially-varying-coefficients-coefficients-varying-wrt-covariate-value}}

Influence from covariates can vary depending on covariate value. For
example, distance to railway station might have a strong impact on
housing price if the distance is small, while it might be weak if the
distance is large. To capture such an effect, the resf function
estimates coefficients varying with respect to covariate value. I call
such coefficients non-spatially varying coefficients (NVCs). If
nvc=TRUE, the resf function estimates the following model considering
NSVs and residual spatial dependence:
\[y_i=\sum^K_{k=1}x_{i,k}\beta_{i,k}+f_{MC}(s_i)+\epsilon_i, \hspace{0.5cm}
\beta_{i,k}=b_k + f(x_{i,k}), \hspace{0.5cm} \epsilon_i \sim N(0, \sigma^2),\]
where \(f(x_{i,k})\) is a smooth function of \(x_{i,k}\) capturing the
non-spatial influence. Here is a code estimating a spatial NVC model
(with selection of constant or NVC):

\begin{Shaded}
\begin{Highlighting}[]
\NormalTok{res   }\OtherTok{\textless{}{-}} \FunctionTok{resf}\NormalTok{(}\AttributeTok{y =}\NormalTok{ y, }\AttributeTok{x =}\NormalTok{ x, }\AttributeTok{meig =}\NormalTok{ meig, }\AttributeTok{nvc=}\ConstantTok{TRUE}\NormalTok{)}
\NormalTok{res}
\end{Highlighting}
\end{Shaded}

\begin{verbatim}
## Call:
## resf(y = y, x = x, nvc = TRUE, meig = meig)
## 
## ----Non-spatially varying coefficients on x (summary) ----
## 
## Coefficients:
##    Intercept          CRIM               ZN              INDUS        
##  Min.   :25.41   Min.   :-0.1822   Min.   :0.02042   Min.   :-0.2119  
##  1st Qu.:25.41   1st Qu.:-0.1822   1st Qu.:0.02042   1st Qu.:-0.2119  
##  Median :25.41   Median :-0.1822   Median :0.02042   Median :-0.2119  
##  Mean   :25.41   Mean   :-0.1822   Mean   :0.02042   Mean   :-0.2119  
##  3rd Qu.:25.41   3rd Qu.:-0.1822   3rd Qu.:0.02042   3rd Qu.:-0.2119  
##  Max.   :25.41   Max.   :-0.1822   Max.   :0.02042   Max.   :-0.2119  
##       CHAS            NOX               RM                AGE          
##  Min.   :1.375   Min.   :-0.463   Min.   :-0.78043   Min.   :-0.06742  
##  1st Qu.:1.375   1st Qu.: 6.083   1st Qu.:-0.40834   1st Qu.:-0.06742  
##  Median :1.375   Median : 7.792   Median :-0.16098   Median :-0.06742  
##  Mean   :1.375   Mean   : 7.074   Mean   : 0.03975   Mean   :-0.06742  
##  3rd Qu.:1.375   3rd Qu.: 8.654   3rd Qu.: 0.19417   3rd Qu.:-0.06742  
##  Max.   :1.375   Max.   :11.517   Max.   : 2.49406   Max.   :-0.06742  
## 
## Statistical significance:
##                         Intercept CRIM  ZN INDUS CHAS NOX  RM AGE
## Not significant                 0    0 506     0    0 506 472   0
## Significant (10% level)         0    0   0     0  506   0   7   0
## Significant ( 5% level)         0    0   0     0    0   0  10   0
## Significant ( 1% level)       506  506   0   506    0   0  17 506
## 
## ----Variance parameter------------------------
## 
## Spatial effects (residuals):
##                      (Intercept)
## random_SD              3.6981527
## Moran.I/max(Moran.I)   0.4490228
## 
## Non-spatial effects (coefficients on x):
##           CRIM ZN INDUS CHAS      NOX        RM AGE
## random_SD    0  0     0    0 1.850518 0.2459548   0
## 
## ----Error statistics--------------------------
##                      stat
## resid_SE        3.7949128
## adjR2(cond)     0.8271073
## rlogLik     -1478.6128728
## AIC          2983.2257457
## BIC          3038.1707224
## 
## NULL model: lm( y ~ x )
##    (r)loglik: -1612.825 ( AIC: 3243.65,  BIC: 3281.689 )
## 
## Note: AIC and BIC are based on the restricted/marginal likelihood.
##       Use method="ml" for comparison of models with different fixed effects (x)
\end{verbatim}

By default, this function selects constant or NVC through BIC
minimization. ``Non-spatially varying coefficients'' in the ``Variance
parameter'' section summarizes the estimated standard errors of the
NVCs. Based on the result, coefficients on \{NOX, RM\} are NVCs, and
coefficients on the others are constants. The NVC on RM, which is the
6-th covariate, is plotted as below. The solid line in the panel denotes
the estimated NVC, and the gray area denotes the 95\% confidence
interval. This plot shows that RM is positively statistically
significant only if RM is large.

\begin{Shaded}
\begin{Highlighting}[]
\FunctionTok{plot\_n}\NormalTok{(res,}\DecValTok{6}\NormalTok{)}
\end{Highlighting}
\end{Shaded}

\includegraphics{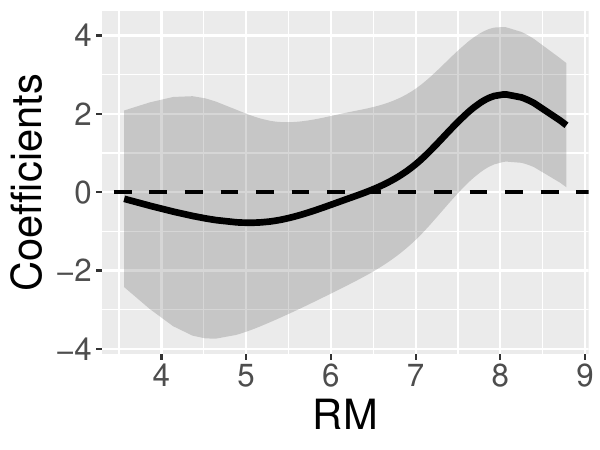}

The NVC can also be spatially plotted as below:

\begin{Shaded}
\begin{Highlighting}[]
\FunctionTok{plot\_s}\NormalTok{(res,}\DecValTok{6}\NormalTok{)}
\end{Highlighting}
\end{Shaded}

\includegraphics{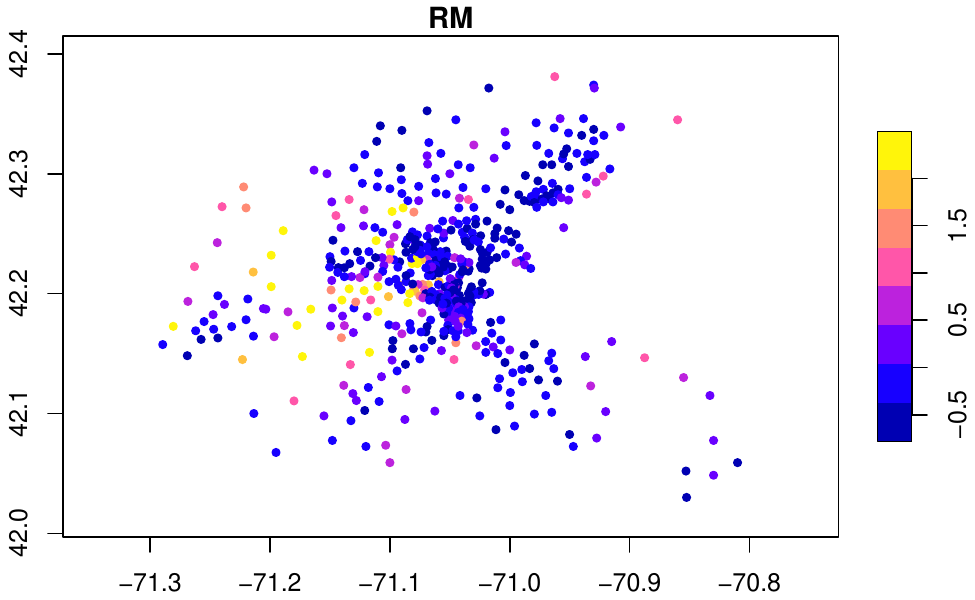}

On the other hand, the residual spatial process \(f_{MC}(s_i)\) is
plotted as

\begin{Shaded}
\begin{Highlighting}[]
\FunctionTok{plot\_s}\NormalTok{(res)}
\end{Highlighting}
\end{Shaded}

\includegraphics{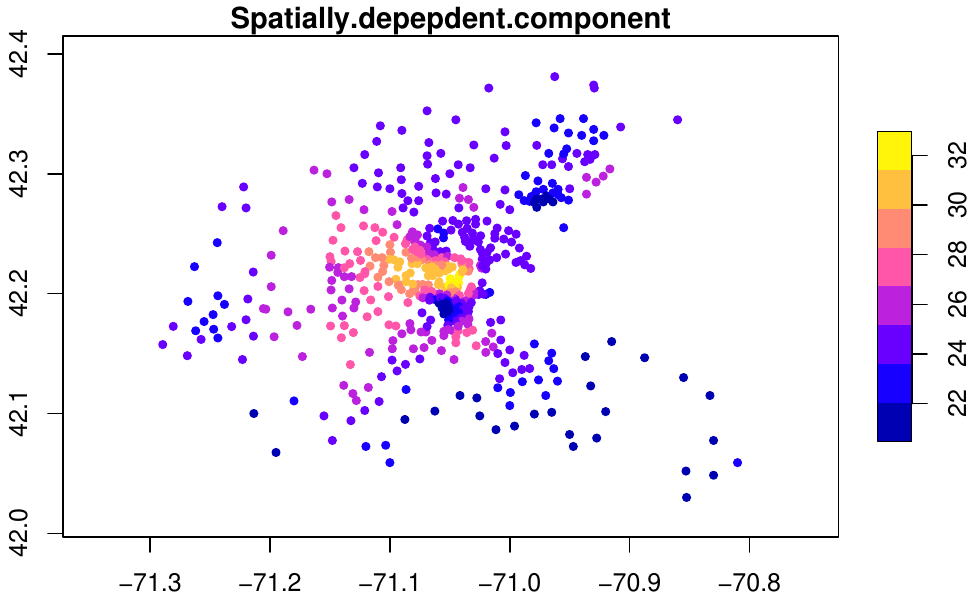}

Sometimes, the user may wish to assume NVCs only on the first three
covariates and constant coefficients on the others. The following code
estimates such a model:

\begin{Shaded}
\begin{Highlighting}[]
\NormalTok{res   }\OtherTok{\textless{}{-}} \FunctionTok{resf}\NormalTok{(}\AttributeTok{y =}\NormalTok{ y, }\AttributeTok{x =}\NormalTok{ x, }\AttributeTok{meig =}\NormalTok{ meig, }\AttributeTok{nvc=}\ConstantTok{TRUE}\NormalTok{, }\AttributeTok{nvc\_sel=}\DecValTok{1}\SpecialCharTok{:}\DecValTok{3}\NormalTok{)}
\end{Highlighting}
\end{Shaded}

\hypertarget{models-with-spatially-varying-coefficients}{%
\subsubsection{Models with spatially varying
coefficients}\label{models-with-spatially-varying-coefficients}}

This package implements an ME-based spatially varying coefficient
(M-SVC) model (Murakami et al., 2017), which is formulated as
\[y_i=\sum^K_{k=1}x_{i,k}\beta_{i,k}+f_{MC}(s_i)+\epsilon_i, \hspace{0.5cm}
\beta_{i,k}=b_k + f_{MC,k}(s_i), \hspace{0.5cm} \epsilon_i \sim N(0, \sigma^2),\]
This model defines the k-th coefficient at site i by \(\beta_{i,k}\)=
{[}constant mean \(b_k\){]} + {[}spatially varying component
\(f_{MC,k}(s_i)\){]}. Geographically weighted regression (GWR) is known
as another SVC estimation approach. Major advantages of the M-SVC
modeling approach over GWR are as follows:

\begin{itemize}
\tightlist
\item
  The M-SVC model estimates the spatial scale (or MC value) of each SVC,
  while the classical GWR assumes a common scale across SVCs.
\item
  The M-SVC model can assume SVCs on some covariates and constant
  coefficients on the others. This is achieved by simply assuming
  \(\beta_{i,k}=b_k\)
\item
  This model is faster and available for very large samples. In
  addition, the model is free from memory limitations if the besf\_vc
  function is used (see Section 4).
\item
  Model selection (i.e., constant coefficient or SVC) is implemented
  without losing its computational efficiency.
\end{itemize}

Here is a sample code estimating an SVC model without coefficient type
selection. In the code, x specifies covariates assuming SVCs, while
xconst specifies covariates assuming constant coefficients. If x\_sel =
FALSE, the types of coefficients on x are fixed.

\begin{Shaded}
\begin{Highlighting}[]
\NormalTok{y         }\OtherTok{\textless{}{-}}\NormalTok{ boston.c[, }\StringTok{"CMEDV"}\NormalTok{]}
\NormalTok{x       }\OtherTok{\textless{}{-}}\NormalTok{ boston.c[,}\FunctionTok{c}\NormalTok{(}\StringTok{"CRIM"}\NormalTok{, }\StringTok{"AGE"}\NormalTok{)]}
\NormalTok{xconst  }\OtherTok{\textless{}{-}}\NormalTok{ boston.c[,}\FunctionTok{c}\NormalTok{(}\StringTok{"ZN"}\NormalTok{,}\StringTok{"DIS"}\NormalTok{,}\StringTok{"RAD"}\NormalTok{,}\StringTok{"NOX"}\NormalTok{,  }\StringTok{"TAX"}\NormalTok{,}\StringTok{"RM"}\NormalTok{, }\StringTok{"PTRATIO"}\NormalTok{, }\StringTok{"B"}\NormalTok{)]}
\NormalTok{coords  }\OtherTok{\textless{}{-}}\NormalTok{ boston.c[,}\FunctionTok{c}\NormalTok{(}\StringTok{"LON"}\NormalTok{,}\StringTok{"LAT"}\NormalTok{)]}
\NormalTok{meig      }\OtherTok{\textless{}{-}} \FunctionTok{meigen}\NormalTok{(}\AttributeTok{coords=}\NormalTok{coords)}
\NormalTok{res     }\OtherTok{\textless{}{-}} \FunctionTok{resf\_vc}\NormalTok{(}\AttributeTok{y=}\NormalTok{y,}\AttributeTok{x=}\NormalTok{x,}\AttributeTok{xconst=}\NormalTok{xconst,}\AttributeTok{meig=}\NormalTok{meig, }\AttributeTok{x\_sel =} \ConstantTok{FALSE}\NormalTok{ )}
\end{Highlighting}
\end{Shaded}

\begin{verbatim}
## [1] "-------  Iteration 1  -------"
## [1] "1/3"
## [1] "2/3"
## [1] "3/3"
## [1] "BIC: 3120.605"
## [1] "-------  Iteration 2  -------"
## [1] "1/3"
## [1] "2/3"
## [1] "3/3"
## [1] "BIC: 3114.252"
## [1] "-------  Iteration 3  -------"
## [1] "1/3"
## [1] "2/3"
## [1] "3/3"
## [1] "BIC: 3114.139"
## [1] "-------  Iteration 4  -------"
## [1] "1/3"
## [1] "2/3"
## [1] "3/3"
## [1] "BIC: 3114.138"
\end{verbatim}

\begin{Shaded}
\begin{Highlighting}[]
\NormalTok{res}
\end{Highlighting}
\end{Shaded}

\begin{verbatim}
## Call:
## resf_vc(y = y, x = x, xconst = xconst, x_sel = FALSE, meig = meig)
## 
## ----Spatially varying coefficients on x (summary)----
## 
## Coefficient estimates:
##   (Intercept)         CRIM               AGE          
##  Min.   :12.03   Min.   :-3.29294   Min.   :-0.14986  
##  1st Qu.:13.99   1st Qu.:-0.19941   1st Qu.:-0.08377  
##  Median :15.06   Median : 0.04993   Median :-0.06780  
##  Mean   :15.70   Mean   : 0.05902   Mean   :-0.06582  
##  3rd Qu.:17.31   3rd Qu.: 0.36587   3rd Qu.:-0.04710  
##  Max.   :20.46   Max.   : 1.83866   Max.   : 0.04298  
## 
## Statistical significance:
##                         Intercept CRIM AGE
## Not significant                 0  416 147
## Significant (10% level)         0   27  40
## Significant ( 5% level)       190   17  99
## Significant ( 1% level)       316   46 220
## 
## ----Constant coefficients on xconst----------------------------
##             Estimate          SE   t_value      p_value
## ZN        0.03202068 0.013219003  2.422322 1.582817e-02
## DIS      -1.47514930 0.334360238 -4.411856 1.292875e-05
## RAD       0.36064288 0.090818317  3.971037 8.368693e-05
## NOX     -36.21088316 5.134427150 -7.052565 6.925571e-12
## TAX      -0.01242296 0.003502523 -3.546862 4.320840e-04
## RM        6.49212566 0.326197980 19.902409 0.000000e+00
## PTRATIO  -0.52573980 0.151594626 -3.468064 5.762765e-04
## B         0.02091202 0.003094117  6.758638 4.477529e-11
## 
## ----Variance parameters----------------------------------
## 
## Spatial effects (coefficients on x):
##                      (Intercept)       CRIM        AGE
## random_SD              3.9039832 1.59443322 0.05746111
## Moran.I/max(Moran.I)   0.6627375 0.04502003 0.06267778
## 
## ----Error statistics-------------------------------------
##                      stat
## resid_SE        3.6706778
## adjR2(cond)     0.8375658
## rlogLik     -1501.0302460
## AIC          3038.0604921
## BIC          3114.1381521
## 
## NULL model: lm( y ~ x + xconst )
##    (r)loglik: -1551.857 ( AIC: 3127.715,  BIC: 3178.433 )
## 
## Note: AIC and BIC are based on the restricted/marginal likelihood.
##       Use method="ml" for comparison of models with different fixed effects (x and xconst)
\end{verbatim}

Estimated SVCs can be plotted as

\begin{Shaded}
\begin{Highlighting}[]
\FunctionTok{plot\_s}\NormalTok{(res,}\DecValTok{0}\NormalTok{) }\CommentTok{\# Spatially varying intercept}
\end{Highlighting}
\end{Shaded}

\includegraphics{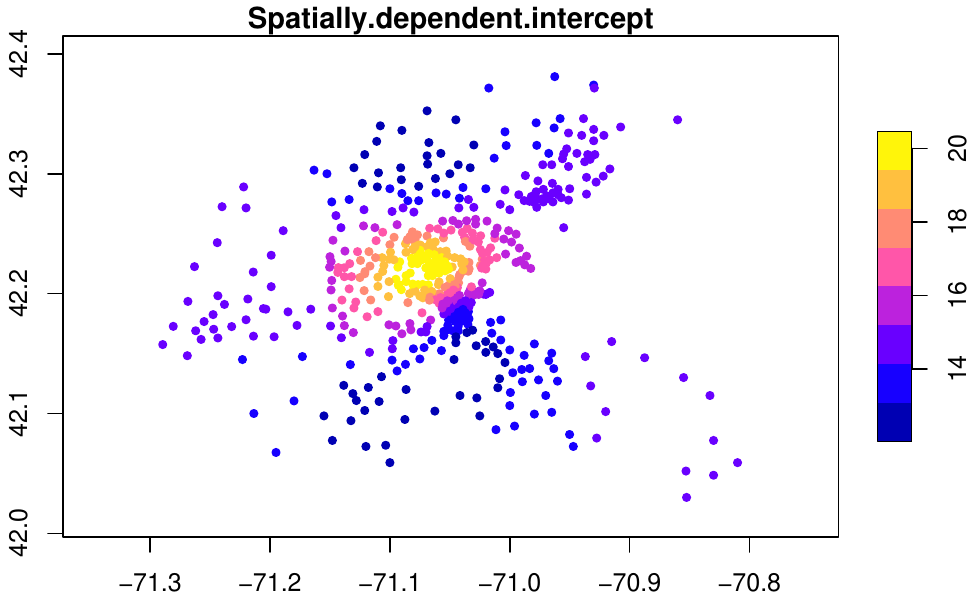}

\begin{Shaded}
\begin{Highlighting}[]
\FunctionTok{plot\_s}\NormalTok{(res,}\DecValTok{1}\NormalTok{) }\CommentTok{\# 1st SVC}
\end{Highlighting}
\end{Shaded}

\includegraphics{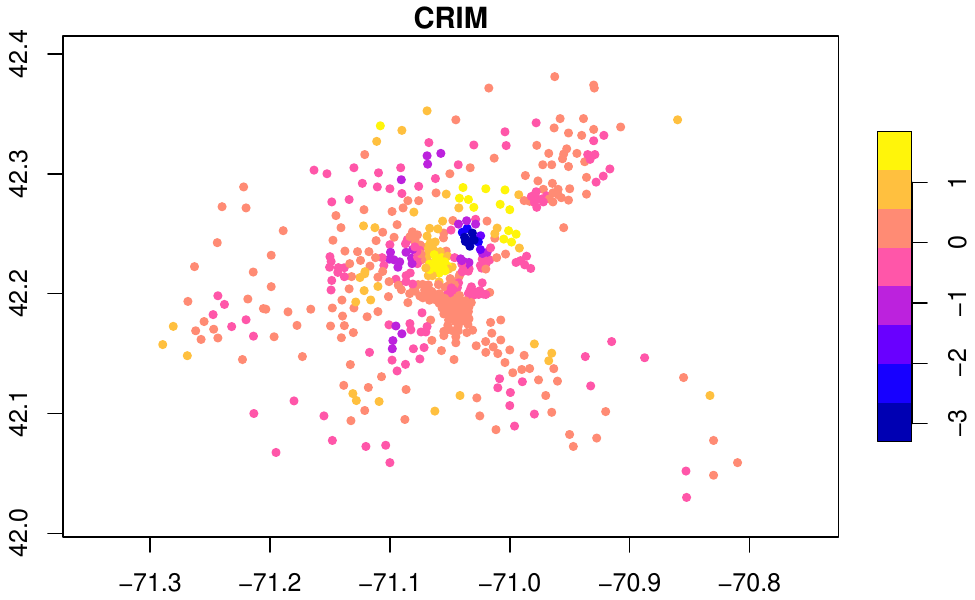}

\begin{Shaded}
\begin{Highlighting}[]
\FunctionTok{plot\_s}\NormalTok{(res,}\DecValTok{2}\NormalTok{) }\CommentTok{\# 2nd SVC}
\end{Highlighting}
\end{Shaded}

\includegraphics{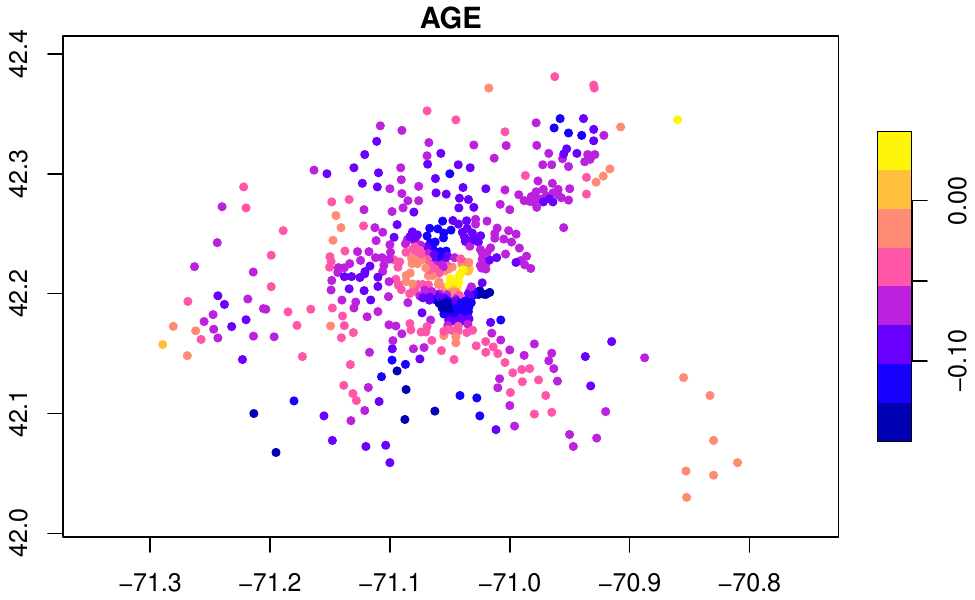}

On the other hand, by default, the resf\_vc function selects constant or
SVCs through a BIC minimization (i.e., x\_sel=TRUE by default). Here is
a code:

\begin{Shaded}
\begin{Highlighting}[]
\NormalTok{res     }\OtherTok{\textless{}{-}} \FunctionTok{resf\_vc}\NormalTok{(}\AttributeTok{y=}\NormalTok{y,}\AttributeTok{x=}\NormalTok{x,}\AttributeTok{xconst=}\NormalTok{xconst,}\AttributeTok{meig=}\NormalTok{meig )}
\end{Highlighting}
\end{Shaded}

\hypertarget{models-with-spatially-and-non-spatially-varying-coefficients}{%
\subsubsection{Models with spatially and non-spatially varying
coefficients}\label{models-with-spatially-and-non-spatially-varying-coefficients}}

The spatially and non-spatially varying coefficient (SNVC) model is
defined as
\[y_i=\sum^K_{k=1}x_{i,k}\beta_{i,k}+f_{MC}(s_i)+\epsilon_i, \hspace{0.5cm}
\beta_{i,k}=b_k + f_{MC,k}(s_i)+f(x_{i,k}), \hspace{0.5cm} \epsilon_i \sim N(0, \sigma^2),\]
This model defines the k-th coefficient as \(\beta_{i,k}\)= {[}constant
mean \(b_k\){]} + {[}spatially varying component \(f_{MC,k}(s_i)\){]} +
{[}non-spatially varying component \(f(x_{i,k})\){]}. Murakami and
Griffith (2020) showed that, unlike SVC models that tend to be unstable
owing to spurious correlation among SVCs (see Wheeler and Tiefelsdorf,
2005), this SNVC model is stable and quite robust against spurious
correlations. Therefore, I recommend using the SNVC model, even if the
purpose of the analysis is estimating SVCs.

An SNVC model is estimated by specifying x\_nvc = TRUE in the resf\_vc
function as follows:

\begin{Shaded}
\begin{Highlighting}[]
\NormalTok{res   }\OtherTok{\textless{}{-}} \FunctionTok{resf\_vc}\NormalTok{(}\AttributeTok{y=}\NormalTok{y,}\AttributeTok{x=}\NormalTok{x,}\AttributeTok{xconst=}\NormalTok{xconst,}\AttributeTok{meig=}\NormalTok{meig, }\AttributeTok{x\_nvc =}\ConstantTok{TRUE}\NormalTok{)}
\end{Highlighting}
\end{Shaded}

This model assumes SNVC on x and constant coefficients on xconst. By
default, the coefficient type (SNVC, SVC, NVC, or constant) on x is
selected.

It is also possible to assume SNVCs on x and NVCs on xcnost by
specifying xconst\_nvc = TRUE as follows:

\begin{Shaded}
\begin{Highlighting}[]
\NormalTok{res   }\OtherTok{\textless{}{-}} \FunctionTok{resf\_vc}\NormalTok{(}\AttributeTok{y=}\NormalTok{y,}\AttributeTok{x=}\NormalTok{x,}\AttributeTok{xconst=}\NormalTok{xconst,}\AttributeTok{meig=}\NormalTok{meig, }\AttributeTok{x\_nvc =}\ConstantTok{TRUE}\NormalTok{, }\AttributeTok{xconst\_nvc=}\ConstantTok{TRUE}\NormalTok{)}
\end{Highlighting}
\end{Shaded}

\begin{verbatim}
## [1] "-------  Iteration 1  -------"
## [1] "1/13"
## [1] "2/13"
## [1] "3/13"
## [1] "4/13"
## [1] "5/13"
## [1] "7/13"
## [1] "8/13"
## [1] "9/13"
## [1] "10/13"
## [1] "11/13"
## [1] "12/13"
## [1] "13/13"
## [1] "BIC: 3023.362"
## [1] "-------  Iteration 2  -------"
## [1] "1/13"
## [1] "2/13"
## [1] "3/13"
## [1] "4/13"
## [1] "5/13"
## [1] "7/13"
## [1] "8/13"
## [1] "9/13"
## [1] "10/13"
## [1] "11/13"
## [1] "12/13"
## [1] "13/13"
## [1] "BIC: 3013.007"
## [1] "-------  Iteration 3  -------"
## [1] "1/13"
## [1] "2/13"
## [1] "3/13"
## [1] "4/13"
## [1] "5/13"
## [1] "7/13"
## [1] "8/13"
## [1] "9/13"
## [1] "10/13"
## [1] "11/13"
## [1] "12/13"
## [1] "13/13"
## [1] "BIC: 3012.859"
## [1] "-------  Iteration 4  -------"
## [1] "1/13"
## [1] "2/13"
## [1] "3/13"
## [1] "4/13"
## [1] "5/13"
## [1] "7/13"
## [1] "8/13"
## [1] "9/13"
## [1] "10/13"
## [1] "11/13"
## [1] "12/13"
## [1] "13/13"
## [1] "BIC: 3012.857"
## [1] "-------  Iteration 5  -------"
## [1] "1/13"
## [1] "2/13"
## [1] "3/13"
## [1] "4/13"
## [1] "5/13"
## [1] "7/13"
## [1] "8/13"
## [1] "9/13"
## [1] "10/13"
## [1] "11/13"
## [1] "12/13"
## [1] "13/13"
## [1] "BIC: 3012.857"
\end{verbatim}

By default, the coefficient type (SNVC, SVC, NVC, or constant) on x and
those (NVC or const) on xconst are selected. The estimated SNVCs are
plotted as follows:

\begin{Shaded}
\begin{Highlighting}[]
\FunctionTok{plot\_s}\NormalTok{(res,}\DecValTok{0}\NormalTok{)           }\CommentTok{\# Spatially varying intercept}
\end{Highlighting}
\end{Shaded}

\includegraphics{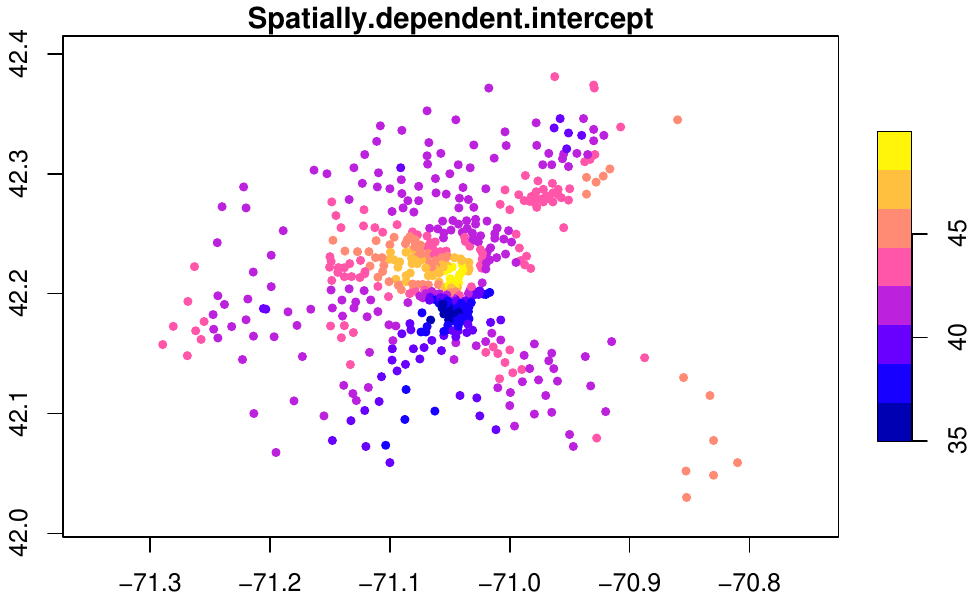}

\begin{Shaded}
\begin{Highlighting}[]
\FunctionTok{plot\_s}\NormalTok{(res,}\DecValTok{1}\NormalTok{)           }\CommentTok{\# SNVC on x[,1]}
\end{Highlighting}
\end{Shaded}

\includegraphics{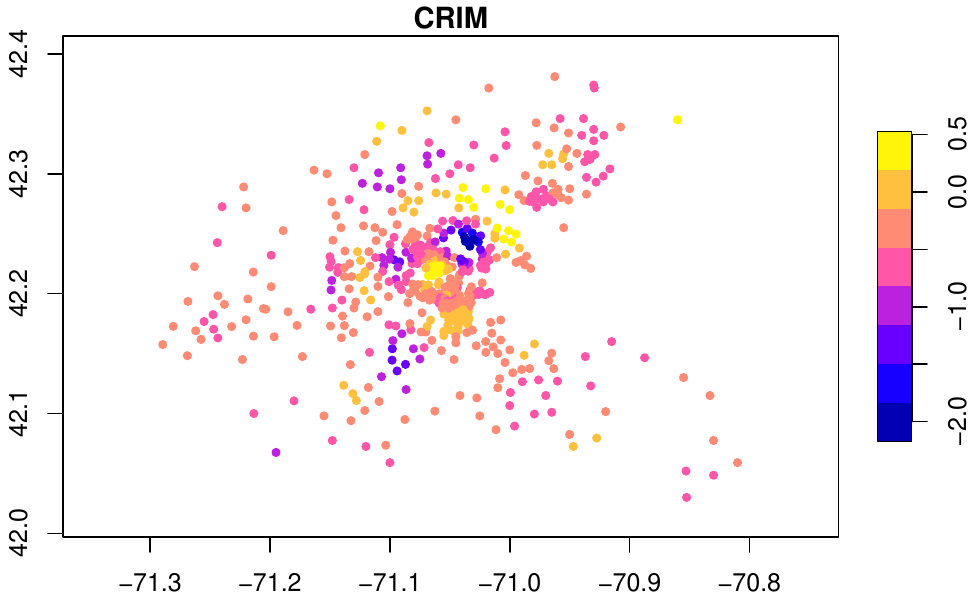}

NVCs on xconst is plotted by specifying xtype=``xconst'' in the plot\_n
function, as below. The solid line denotes the estimated NVC, and the
gray area denotes the 95\% confidence interval:

\begin{Shaded}
\begin{Highlighting}[]
\FunctionTok{plot\_n}\NormalTok{(res,}\DecValTok{4}\NormalTok{,}\AttributeTok{xtype=}\StringTok{"xconst"}\NormalTok{)}\CommentTok{\#NVC on xconst[,4]}
\end{Highlighting}
\end{Shaded}

\includegraphics{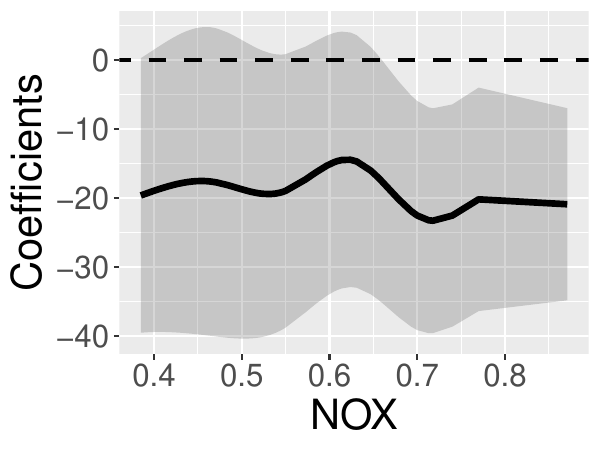}

\begin{Shaded}
\begin{Highlighting}[]
\FunctionTok{plot\_n}\NormalTok{(res,}\DecValTok{6}\NormalTok{,}\AttributeTok{xtype=}\StringTok{"xconst"}\NormalTok{)}\CommentTok{\#NVC on xconst[,6]}
\end{Highlighting}
\end{Shaded}

\includegraphics{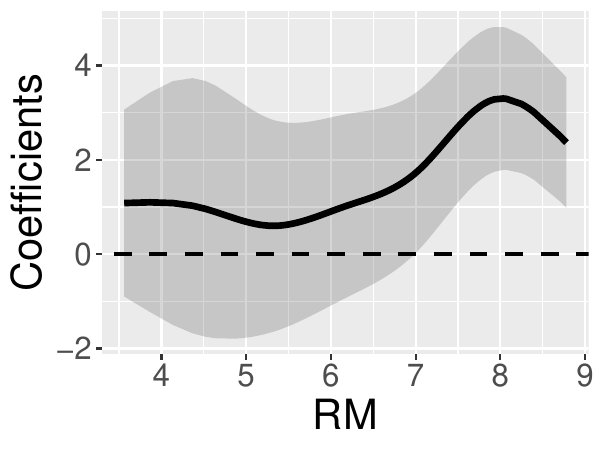}

These NVCs can also be plotted spatially as follows:

\begin{Shaded}
\begin{Highlighting}[]
\FunctionTok{plot\_s}\NormalTok{(res,}\DecValTok{4}\NormalTok{,}\AttributeTok{xtype=}\StringTok{"xconst"}\NormalTok{)}\CommentTok{\#Spatial plot of NVC on xconst[,4]}
\end{Highlighting}
\end{Shaded}

\includegraphics{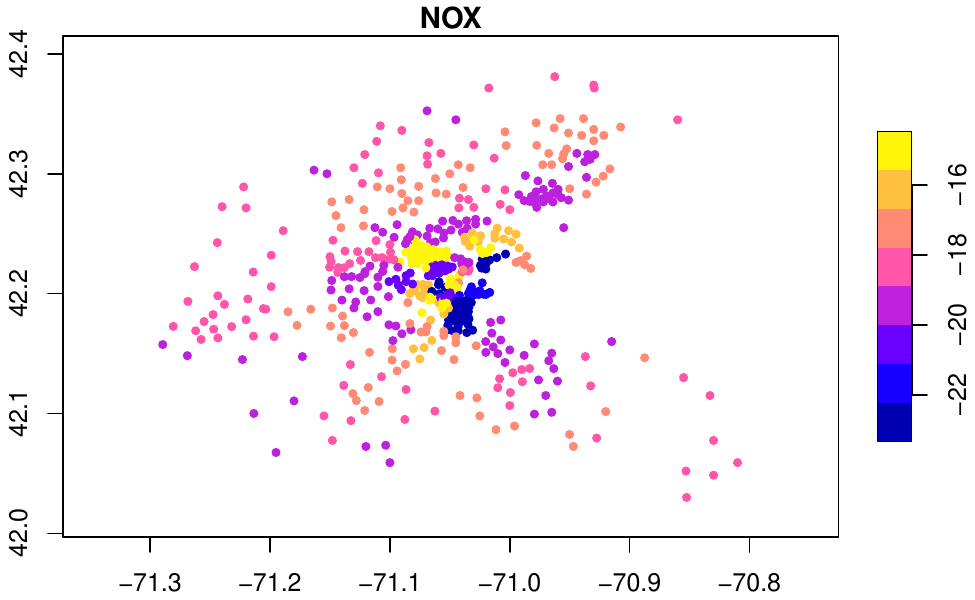}

\begin{Shaded}
\begin{Highlighting}[]
\FunctionTok{plot\_s}\NormalTok{(res,}\DecValTok{6}\NormalTok{,}\AttributeTok{xtype=}\StringTok{"xconst"}\NormalTok{)}\CommentTok{\#Spatial plot of NVC on xconst[,6]}
\end{Highlighting}
\end{Shaded}

\includegraphics{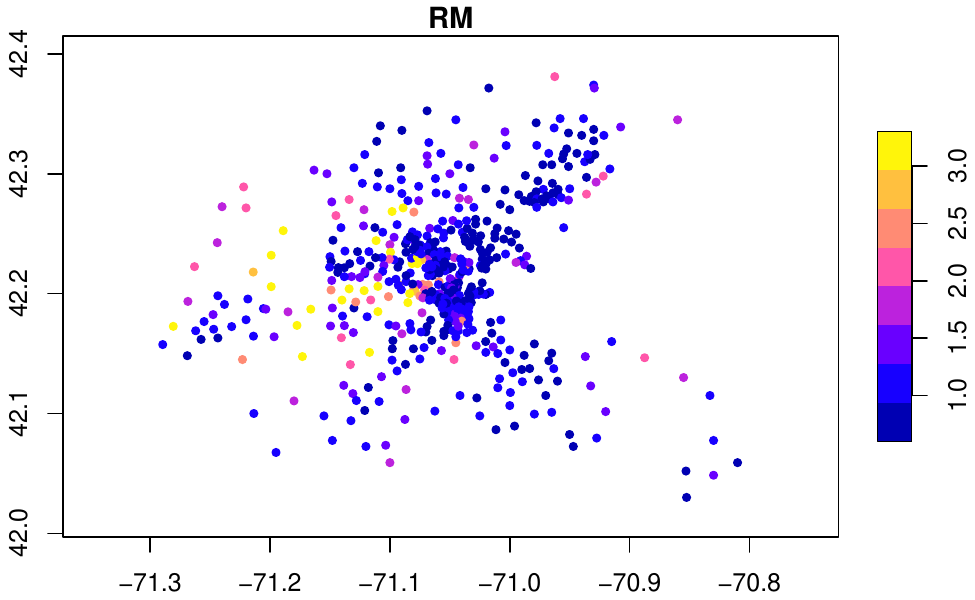}

\hypertarget{models-with-group-effects}{%
\subsubsection{Models with group
effects}\label{models-with-group-effects}}

\hypertarget{outline}{%
\paragraph{Outline}\label{outline}}

Two group effects are available in this package:

\begin{enumerate}
\def\labelenumi{\arabic{enumi}.}
\tightlist
\item
  Spatially dependent group effects. Spatial dependence among groups is
  modeled instead of modeling spatial dependence among individuals.
\item
  Spatially independent group effects assuming independence across
  groups (usual group effects)
\end{enumerate}

They are estimated in the resf and resf\_vc functions. When considering
both these effects, the resf function estimates the following model (if
no NVC is assumed):
\[y_i=\sum^K_{k=1}x_{i,k}\beta_k+f_{MC}(g_{I(0)})+\sum^H_{h=1}\gamma(g_{I(h)})+\epsilon_i, \hspace{0.5cm}\epsilon_i \sim N(0, \sigma^2),\]
where \(g_{I(0)},g_{I(1)},\ldots,g_{I(H)}\) represent group variables.
\(f_{MC}(g_{I(0)})\) denotes spatially dependent group effects, while
\(\gamma(g_{I(h)})\) denotes spatially independent group effects for the
h-th group variable. On the other hand, the resf\_vc function can
estimate the following model considering these two effects (again, no
NVC is assumed):
\[y_i=\sum^K_{k=1}x_{i,k}\beta_{i,k}+f_{MC}(g_{I(0)})+\sum^H_{h=1}\gamma(g_{I(h)})+\epsilon_i, \hspace{0.5cm}
\beta_{i,k}=b_k + f_{MC,k}(g_{i(0)}), \hspace{0.5cm} \epsilon_i \sim N(0, \sigma^2),\]
Below, multilevel modeling, small area estimation, and panel data
analysis are demonstrated.

\hypertarget{multilevel-model}{%
\paragraph{Multilevel model}\label{multilevel-model}}

Data often have a multilevel structure. For example, the school
achievement of individual students changes depending on the class and
school. A condominium unit price depends, not only on unit attributes,
but also on building attributes. Multilevel modeling is required to
explicitly consider the multilevel structure behind data and perform
spatial regressions.

This section demonstrates the modeling considering the two group effects
using the resf function. The data used are the Boston housing datasets
that consist of 506 samples in 92 towns, which are regarded as groups.
To model spatially dependent group effects, Moran eigenvectors are
defined by groups. This is done by specifying s\_id in the meigen
function using a group variable, which is the town name (TOWNNO), in
this case, as follows:

\begin{Shaded}
\begin{Highlighting}[]
\NormalTok{xgroup}\OtherTok{\textless{}{-}}\NormalTok{ boston.c[,}\StringTok{"TOWNNO"}\NormalTok{]}
\NormalTok{coords}\OtherTok{\textless{}{-}}\NormalTok{ boston.c[,}\FunctionTok{c}\NormalTok{(}\StringTok{"LON"}\NormalTok{,}\StringTok{"LAT"}\NormalTok{)]}
\NormalTok{meig\_g}\OtherTok{\textless{}{-}} \FunctionTok{meigen}\NormalTok{(}\AttributeTok{coords=}\NormalTok{coords, }\AttributeTok{s\_id=}\NormalTok{xgroup)}
\end{Highlighting}
\end{Shaded}

When additionally estimating spatially independent group effects, the
user needs to specify xgroup in the resf function by one or more group
variables, as follows:

\begin{Shaded}
\begin{Highlighting}[]
\NormalTok{x       }\OtherTok{\textless{}{-}}\NormalTok{ boston.c[,}\FunctionTok{c}\NormalTok{(}\StringTok{"CRIM"}\NormalTok{,}\StringTok{"ZN"}\NormalTok{,}\StringTok{"INDUS"}\NormalTok{, }\StringTok{"CHAS"}\NormalTok{, }\StringTok{"NOX"}\NormalTok{,}\StringTok{"RM"}\NormalTok{, }\StringTok{"AGE"}\NormalTok{)]}
\NormalTok{res   }\OtherTok{\textless{}{-}} \FunctionTok{resf}\NormalTok{(}\AttributeTok{y =}\NormalTok{ y, }\AttributeTok{x =}\NormalTok{ x, }\AttributeTok{meig =}\NormalTok{ meig\_g, }\AttributeTok{xgroup =}\NormalTok{ xgroup)}
\NormalTok{res}
\end{Highlighting}
\end{Shaded}

\begin{verbatim}
## Call:
## resf(y = y, x = x, xgroup = xgroup, meig = meig_g)
## 
## ----Coefficients------------------------------
##                 Estimate         SE    t_value      p_value
## (Intercept)  -0.81545944 3.23135854 -0.2523581 8.008871e-01
## CRIM         -0.04596392 0.02505503 -1.8345188 6.728064e-02
## ZN            0.03285021 0.02313784  1.4197611 1.564153e-01
## INDUS         0.03549188 0.11980486  0.2962474 7.671869e-01
## CHAS         -0.62561231 0.72381491 -0.8643264 3.878995e-01
## NOX         -26.38632671 3.88238119 -6.7964286 3.668488e-11
## RM            6.30273567 0.29409796 21.4307357 0.000000e+00
## AGE          -0.06730232 0.01048068 -6.4215611 3.637544e-10
## 
## ----Variance parameter------------------------
## 
## Spatial effects (residuals):
##                      (Intercept)
## random_SD               5.074794
## Moran.I/max(Moran.I)    0.812936
## 
## Group effects:
##           xgroup
## ramdom_SD 4.4404
## 
## ----Error statistics--------------------------
##                      stat
## resid_SE        3.2429178
## adjR2(cond)     0.8740022
## rlogLik     -1465.8457138
## AIC          2955.6914276
## BIC          3006.4098677
## 
## NULL model: lm( y ~ x )
##    (r)loglik: -1612.825 ( AIC: 3243.65,  BIC: 3281.689 )
## 
## Note: AIC and BIC are based on the restricted/marginal likelihood.
##       Use method="ml" for comparison of models with different fixed effects (x)
\end{verbatim}

The estimated independent group effects are extracted as

\begin{Shaded}
\begin{Highlighting}[]
\NormalTok{res}\SpecialCharTok{$}\NormalTok{b\_g[[}\DecValTok{1}\NormalTok{]][}\DecValTok{1}\SpecialCharTok{:}\DecValTok{5}\NormalTok{,]}\CommentTok{\# Estimates in the first 5 groups}
\end{Highlighting}
\end{Shaded}

\begin{verbatim}
##          Estimate       SE   t_value
## xgroup_0 2.165726 2.061093 1.0507657
## xgroup_1 3.747633 1.783543 2.1012294
## xgroup_2 6.544205 1.659184 3.9442318
## xgroup_3 2.431558 1.431325 1.6988163
## xgroup_4 1.036033 1.181672 0.8767521
\end{verbatim}

\hypertarget{small-area-estimation}{%
\paragraph{Small area estimation}\label{small-area-estimation}}

Small area estimation (SAE; Ghosh and Rao, 1994) is a statistical
technique estimating parameters for small areas such as districts and
municipality. SAE is useful for obtaining reliable small area statistics
from noisy data. The resf and resf\_vc functions are available for SEA
(see as explained in Murakami 2020 for further detail).

The Boston housing datasets consist of 506 samples in 92 towns. This
section estimates the standard housing price in the I-th towns by
assuming the following model:
\[y_I=\hat{y}_I+ \epsilon_I, \hspace{0.5cm} \epsilon_I \sim N ( 0, \frac{\sigma^2}{N_I} ) \hspace{0.5cm}\]
where \(\hat{y}_I=\sum^{N_I}_{i=1}{\frac{\hat{y}_i}{N_I}}\). This model
decomposes the observed mean house price \(y_I\) in the I-th town into
the standard price \(\hat{y}_I\) and noise \(\epsilon_I\), which reduces
as the number of samples in the I-th town increases. The standard price
is defined by an aggregate of the predictors \(\hat{y}_i\) by
individuals.

The above equation suggests that, if \(\hat{y}_i\) is predicted using
the resf or resf\_vc function and aggregated into the towns, we can
estimate the standard house price. Here is a sample code for the
individual level prediction:

\begin{Shaded}
\begin{Highlighting}[]
\NormalTok{r\_res }\OtherTok{\textless{}{-}}\FunctionTok{resf}\NormalTok{(}\AttributeTok{y=}\NormalTok{y, }\AttributeTok{x=}\NormalTok{x, }\AttributeTok{meig=}\NormalTok{meig\_g, }\AttributeTok{xgroup=}\NormalTok{xgroup)}
\NormalTok{pred  }\OtherTok{\textless{}{-}}\FunctionTok{predict0}\NormalTok{(r\_res, }\AttributeTok{x0=}\NormalTok{x, }\AttributeTok{meig0=}\NormalTok{meig\_g, }\AttributeTok{xgroup0=}\NormalTok{xgroup)}
\NormalTok{pred}\SpecialCharTok{$}\NormalTok{pred[}\DecValTok{1}\SpecialCharTok{:}\DecValTok{5}\NormalTok{,]}
\end{Highlighting}
\end{Shaded}

\begin{verbatim}
##       pred       xb sf_residual   xgroup
## 1 23.70932 22.71407   -1.170482 2.165726
## 2 24.57615 22.21874   -1.390220 3.747633
## 3 30.58942 28.23201   -1.390220 3.747633
## 4 33.24998 28.19959   -1.493814 6.544205
## 5 33.62206 28.57167   -1.493814 6.544205
\end{verbatim}

As shown above, the predict0 function returns predicted values (pred),
predicted trends (xb), predicted residual spatial components
(sf\_residuals), and predicted group effects (xgroup). Then, these
individual-level variables are aggregated into towns. Here is a code:

\begin{Shaded}
\begin{Highlighting}[]
\NormalTok{adat  }\OtherTok{\textless{}{-}} \FunctionTok{aggregate}\NormalTok{(}\FunctionTok{data.frame}\NormalTok{(y, pred}\SpecialCharTok{$}\NormalTok{pred),}\AttributeTok{by=}\FunctionTok{list}\NormalTok{(xgroup),mean)}
\NormalTok{adat[}\DecValTok{1}\SpecialCharTok{:}\DecValTok{5}\NormalTok{,]}
\end{Highlighting}
\end{Shaded}

\begin{verbatim}
##   Group.1        y     pred       xb sf_residual   xgroup
## 1       0 24.00000 23.70932 22.71407   -1.170482 2.165726
## 2       1 28.15000 27.58279 25.22537   -1.390220 3.747633
## 3       2 32.76667 31.89132 26.84093   -1.493814 6.544205
## 4       3 19.42857 19.36679 18.51187   -1.576641 2.431558
## 5       4 16.71364 16.72781 17.10793   -1.416151 1.036033
\end{verbatim}

The outputs are the predicted standard price (pred), trend (xb),
spatially dependent group effects (sf\_residual), and spatially
independent group effects (xgroup) by town.

To map the result, spatial polygons for the towns are loaded and
combined with our estimates:

\begin{Shaded}
\begin{Highlighting}[]
\FunctionTok{require}\NormalTok{(dplyr)}
\NormalTok{b1           }\OtherTok{\textless{}{-}} \FunctionTok{st\_read}\NormalTok{(}\FunctionTok{system.file}\NormalTok{(}\StringTok{"shapes/boston\_tracts.shp"}\NormalTok{,}\AttributeTok{package=}\StringTok{"spData"}\NormalTok{)[}\DecValTok{1}\NormalTok{])}
\end{Highlighting}
\end{Shaded}

\begin{verbatim}
## Reading layer `boston_tracts' from data source 
##   `C:\Users\dmura\AppData\Local\R\win-library\4.3\spData\shapes\boston_tracts.shp' 
##   using driver `ESRI Shapefile'
## Simple feature collection with 506 features and 36 fields
## Geometry type: POLYGON
## Dimension:     XY
## Bounding box:  xmin: -71.52311 ymin: 42.00305 xmax: -70.63823 ymax: 42.67307
## Geodetic CRS:  NAD27
\end{verbatim}

\begin{Shaded}
\begin{Highlighting}[]
\NormalTok{boston.tr2   }\OtherTok{\textless{}{-}}\NormalTok{ b1 }\SpecialCharTok{\%\textgreater{}\%} \FunctionTok{group\_by}\NormalTok{(TOWNNO) }\SpecialCharTok{\%\textgreater{}\%} \FunctionTok{summarize}\NormalTok{() }\CommentTok{\#dissolve}
\NormalTok{boston.tr2}\SpecialCharTok{$}\NormalTok{id}\OtherTok{\textless{}{-}} \DecValTok{1}\SpecialCharTok{:}\NormalTok{(}\FunctionTok{dim}\NormalTok{(boston.tr2)[}\DecValTok{1}\NormalTok{])}
\NormalTok{boston.tr3   }\OtherTok{\textless{}{-}} \FunctionTok{merge}\NormalTok{(boston.tr2, adat,}\AttributeTok{by.x=}\StringTok{"TOWNNO"}\NormalTok{,}\AttributeTok{by.y=}\StringTok{"Group.1"}\NormalTok{,}\AttributeTok{all.x=}\ConstantTok{TRUE}\NormalTok{)}
\end{Highlighting}
\end{Shaded}

Here are the maps of our estimates. ``y'' denotes the observed mean
prices, and ``pred'' denotes our predicted standard price. While they
are similar, there are some differences in towns with high housing
prices.

\begin{Shaded}
\begin{Highlighting}[]
\NormalTok{boston.tr4  }\OtherTok{\textless{}{-}}\NormalTok{ boston.tr3[}\FunctionTok{order}\NormalTok{(boston.tr3}\SpecialCharTok{$}\NormalTok{id),]}
\FunctionTok{plot}\NormalTok{(boston.tr4[,}\FunctionTok{c}\NormalTok{(}\StringTok{"y"}\NormalTok{,}\StringTok{"pred"}\NormalTok{)], }\AttributeTok{lwd=}\FloatTok{0.3}\NormalTok{,}\AttributeTok{axes=}\ConstantTok{TRUE}\NormalTok{,}\AttributeTok{key.pos=}\DecValTok{4}\NormalTok{)}
\end{Highlighting}
\end{Shaded}

\includegraphics{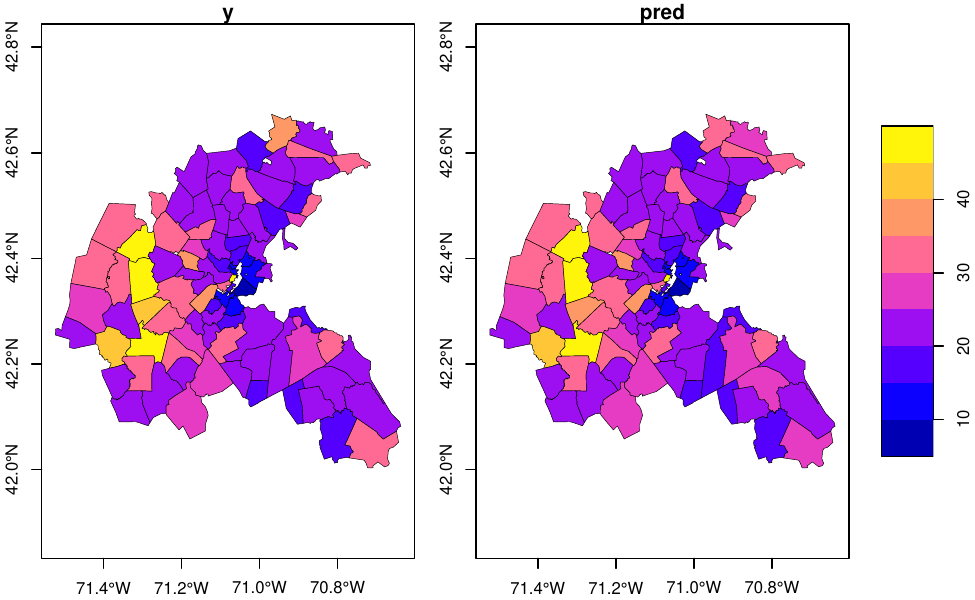}

Here are the elements of the predicted values. The maps below show that
each element explains different things to each other:

\begin{Shaded}
\begin{Highlighting}[]
\FunctionTok{plot}\NormalTok{(boston.tr4[,}\FunctionTok{c}\NormalTok{(}\StringTok{"xgroup"}\NormalTok{,}\StringTok{"sf\_residual"}\NormalTok{)], }\AttributeTok{lwd=}\FloatTok{0.3}\NormalTok{,}\AttributeTok{axes=}\ConstantTok{TRUE}\NormalTok{,}\AttributeTok{key.pos=}\DecValTok{4}\NormalTok{)}
\end{Highlighting}
\end{Shaded}

\includegraphics{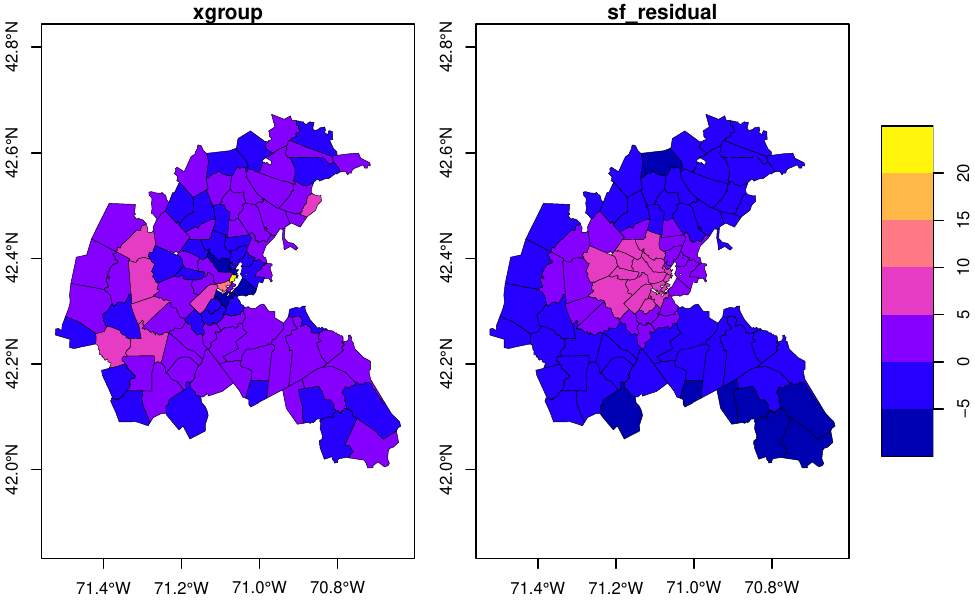}

\begin{Shaded}
\begin{Highlighting}[]
\FunctionTok{plot}\NormalTok{(boston.tr4[,}\StringTok{"xb"}\NormalTok{], }\AttributeTok{lwd=}\FloatTok{0.3}\NormalTok{,}\AttributeTok{axes=}\ConstantTok{TRUE}\NormalTok{,}\AttributeTok{key.pos=}\DecValTok{4}\NormalTok{,}\AttributeTok{nbreaks=}\DecValTok{20}\NormalTok{)}
\end{Highlighting}
\end{Shaded}

\includegraphics{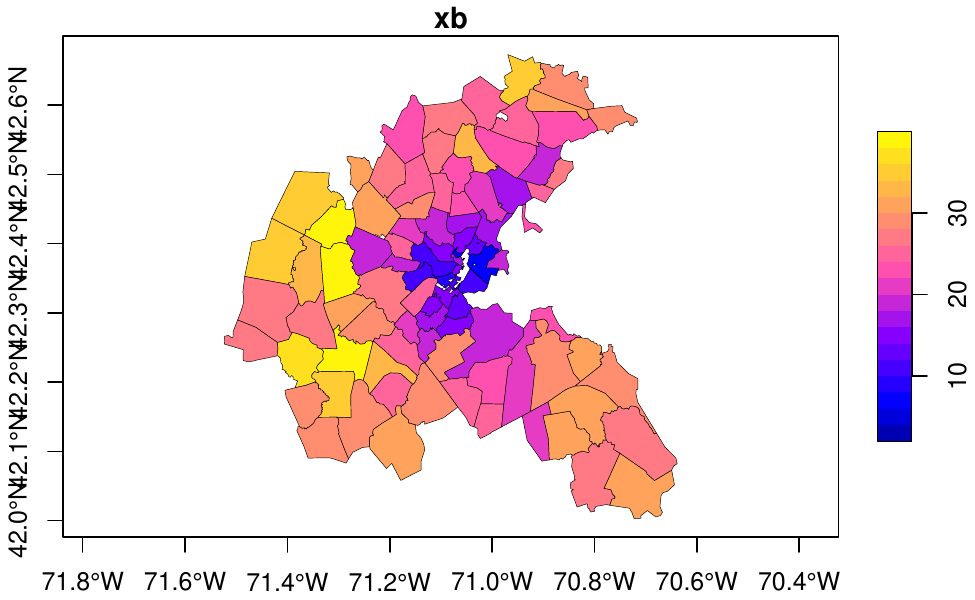}

Note that the resf\_vc function is also available for SVC model-based
SAE. Here is a sample code:

\begin{Shaded}
\begin{Highlighting}[]
\NormalTok{rv\_res  }\OtherTok{\textless{}{-}} \FunctionTok{resf\_vc}\NormalTok{(}\AttributeTok{y=}\NormalTok{y, }\AttributeTok{x=}\NormalTok{x, }\AttributeTok{meig=}\NormalTok{meig\_g, }\AttributeTok{xgroup=}\NormalTok{xgroup, }\AttributeTok{x\_sel=}\ConstantTok{FALSE}\NormalTok{)}
\end{Highlighting}
\end{Shaded}

\begin{verbatim}
## [1] "-------  Iteration 1  -------"
## [1] "1/9"
## [1] "2/9"
## [1] "3/9"
## [1] "4/9"
## [1] "5/9"
## [1] "6/9"
## [1] "7/9"
## [1] "8/9"
## [1] "9/9"
## [1] "BIC: 3088.013"
## [1] "-------  Iteration 2  -------"
## [1] "1/9"
## [1] "2/9"
## [1] "3/9"
## [1] "4/9"
## [1] "5/9"
## [1] "6/9"
## [1] "7/9"
## [1] "8/9"
## [1] "9/9"
## [1] "BIC: 3047.815"
## [1] "-------  Iteration 3  -------"
## [1] "1/9"
## [1] "2/9"
## [1] "3/9"
## [1] "4/9"
## [1] "5/9"
## [1] "6/9"
## [1] "7/9"
## [1] "8/9"
## [1] "9/9"
## [1] "BIC: 3039.75"
## [1] "-------  Iteration 4  -------"
## [1] "1/9"
## [1] "2/9"
## [1] "3/9"
## [1] "4/9"
## [1] "5/9"
## [1] "6/9"
## [1] "7/9"
## [1] "8/9"
## [1] "9/9"
## [1] "BIC: 3039.572"
## [1] "-------  Iteration 5  -------"
## [1] "1/9"
## [1] "2/9"
## [1] "3/9"
## [1] "4/9"
## [1] "5/9"
## [1] "6/9"
## [1] "7/9"
## [1] "8/9"
## [1] "9/9"
## [1] "BIC: 3039.572"
\end{verbatim}

\begin{Shaded}
\begin{Highlighting}[]
\NormalTok{pred\_vc }\OtherTok{\textless{}{-}} \FunctionTok{predict0}\NormalTok{(rv\_res, }\AttributeTok{x0=}\NormalTok{x, }\AttributeTok{meig0=}\NormalTok{meig\_g, }\AttributeTok{xgroup0=}\NormalTok{xgroup)}
\NormalTok{adat\_vc }\OtherTok{\textless{}{-}} \FunctionTok{aggregate}\NormalTok{(}\FunctionTok{data.frame}\NormalTok{(y, pred\_vc}\SpecialCharTok{$}\NormalTok{pred), }\AttributeTok{by=}\FunctionTok{list}\NormalTok{(xgroup), mean)}
\NormalTok{adat\_vc[}\DecValTok{1}\SpecialCharTok{:}\DecValTok{5}\NormalTok{,]}
\end{Highlighting}
\end{Shaded}

\begin{verbatim}
##   Group.1        y     pred       xb sf_residual   xgroup
## 1       0 24.00000 23.67904 23.12338   -1.123777 1.679437
## 2       1 28.15000 27.81258 27.44293   -1.963678 2.333325
## 3       2 32.76667 32.28742 31.09238   -2.547910 3.742953
## 4       3 19.42857 19.25699 18.45332   -2.501985 3.305659
## 5       4 16.71364 16.68361 15.40092   -1.024469 2.307158
\end{verbatim}

\hypertarget{longitudinalpanel-data-analysis}{%
\paragraph{Longitudinal/panel data
analysis}\label{longitudinalpanel-data-analysis}}

The resf and resf\_vc functions are also available for longitudinal or
panel data analysis with/without S(N)VC (see Yu et al., 2020). Although
this section takes resf as an example, resf\_vc function-based panel
analysis is implemented in the same way.

To illustrate this, we use a panel data of 48 US states from 1970 to
1986, which is published in the plm package (Croissant and Millo, 2008).
Because our approach uses spatial coordinates by default, we added
center spatial coordinates (px and py) to the panel data. Here is the
code:

\begin{Shaded}
\begin{Highlighting}[]
\FunctionTok{require}\NormalTok{(plm)}
\FunctionTok{require}\NormalTok{(spData)}

\FunctionTok{data}\NormalTok{(Produc, }\AttributeTok{package =} \StringTok{"plm"}\NormalTok{)}
\FunctionTok{data}\NormalTok{(us\_states)}
\NormalTok{us\_states2 }\OtherTok{\textless{}{-}} \FunctionTok{data.frame}\NormalTok{(us\_states}\SpecialCharTok{$}\NormalTok{GEOID,us\_states}\SpecialCharTok{$}\NormalTok{NAME,}\FunctionTok{st\_coordinates}\NormalTok{(}\FunctionTok{st\_centroid}\NormalTok{(us\_states)))}
\FunctionTok{names}\NormalTok{(us\_states2)[}\DecValTok{3}\SpecialCharTok{:}\DecValTok{4}\NormalTok{]}\OtherTok{\textless{}{-}} \FunctionTok{c}\NormalTok{(}\StringTok{"px"}\NormalTok{,}\StringTok{"py"}\NormalTok{)}
\NormalTok{us\_states3 }\OtherTok{\textless{}{-}}\NormalTok{ us\_states2[}\FunctionTok{order}\NormalTok{(us\_states2[,}\DecValTok{1}\NormalTok{]),][}\SpecialCharTok{{-}}\DecValTok{8}\NormalTok{,]}
\NormalTok{us\_states3}\SpecialCharTok{$}\NormalTok{state}\OtherTok{\textless{}{-}} \FunctionTok{unique}\NormalTok{(Produc[,}\DecValTok{1}\NormalTok{])}
\NormalTok{pdat0      }\OtherTok{\textless{}{-}} \FunctionTok{na.omit}\NormalTok{(}\FunctionTok{merge}\NormalTok{(Produc,us\_states3[,}\FunctionTok{c}\NormalTok{(}\DecValTok{3}\SpecialCharTok{:}\DecValTok{5}\NormalTok{)],}\AttributeTok{by=}\StringTok{"state"}\NormalTok{,}\AttributeTok{all.x=}\ConstantTok{TRUE}\NormalTok{,}\AttributeTok{sort=}\ConstantTok{FALSE}\NormalTok{))}
\NormalTok{pdat       }\OtherTok{\textless{}{-}}\NormalTok{ pdat0[}\FunctionTok{order}\NormalTok{(pdat0}\SpecialCharTok{$}\NormalTok{state,pdat0}\SpecialCharTok{$}\NormalTok{year),]}
\NormalTok{pdat[}\DecValTok{1}\SpecialCharTok{:}\DecValTok{5}\NormalTok{,]}
\end{Highlighting}
\end{Shaded}

\begin{verbatim}
##     state year region     pcap     hwy   water    util       pc   gsp    emp
## 1 ALABAMA 1970      6 15032.67 7325.80 1655.68 6051.20 35793.80 28418 1010.5
## 2 ALABAMA 1971      6 15501.94 7525.94 1721.02 6254.98 37299.91 29375 1021.9
## 3 ALABAMA 1972      6 15972.41 7765.42 1764.75 6442.23 38670.30 31303 1072.3
## 4 ALABAMA 1973      6 16406.26 7907.66 1742.41 6756.19 40084.01 33430 1135.5
## 5 ALABAMA 1974      6 16762.67 8025.52 1734.85 7002.29 42057.31 33749 1169.8
##   unemp        px       py
## 1   4.7 -86.82797 32.78034
## 2   5.2 -86.82797 32.78034
## 3   4.7 -86.82797 32.78034
## 4   3.9 -86.82797 32.78034
## 5   5.5 -86.82797 32.78034
\end{verbatim}

Here are the first five rows of the data:

\begin{Shaded}
\begin{Highlighting}[]
\NormalTok{pdat[}\DecValTok{1}\SpecialCharTok{:}\DecValTok{5}\NormalTok{,]}
\end{Highlighting}
\end{Shaded}

\begin{verbatim}
##     state year region     pcap     hwy   water    util       pc   gsp    emp
## 1 ALABAMA 1970      6 15032.67 7325.80 1655.68 6051.20 35793.80 28418 1010.5
## 2 ALABAMA 1971      6 15501.94 7525.94 1721.02 6254.98 37299.91 29375 1021.9
## 3 ALABAMA 1972      6 15972.41 7765.42 1764.75 6442.23 38670.30 31303 1072.3
## 4 ALABAMA 1973      6 16406.26 7907.66 1742.41 6756.19 40084.01 33430 1135.5
## 5 ALABAMA 1974      6 16762.67 8025.52 1734.85 7002.29 42057.31 33749 1169.8
##   unemp        px       py
## 1   4.7 -86.82797 32.78034
## 2   5.2 -86.82797 32.78034
## 3   4.7 -86.82797 32.78034
## 4   3.9 -86.82797 32.78034
## 5   5.5 -86.82797 32.78034
\end{verbatim}

Following a vignette of the plm package, this section uses logged gross
state product as explained variables (y) and logged public capital stock
(log\_pcap), logged private capital stock (log\_pc), logged labor input
measured by the employment in non-agricultural payrolls (log\_emp), and
unemployment rate (unemp) as covariables.

\begin{Shaded}
\begin{Highlighting}[]
\NormalTok{y     }\OtherTok{\textless{}{-}} \FunctionTok{log}\NormalTok{(pdat}\SpecialCharTok{$}\NormalTok{gsp)}
\NormalTok{x     }\OtherTok{\textless{}{-}} \FunctionTok{data.frame}\NormalTok{(}\AttributeTok{log\_pcap=}\FunctionTok{log}\NormalTok{(pdat}\SpecialCharTok{$}\NormalTok{pcap), }\AttributeTok{log\_pc=}\FunctionTok{log}\NormalTok{(pdat}\SpecialCharTok{$}\NormalTok{pc),}
                    \AttributeTok{log\_emp=}\FunctionTok{log}\NormalTok{(pdat}\SpecialCharTok{$}\NormalTok{emp), }\AttributeTok{unemp=}\NormalTok{pdat}\SpecialCharTok{$}\NormalTok{unemp)}
\end{Highlighting}
\end{Shaded}

Because spatial coordinates are defined by states, Moran eigenvectors
must be extracted by state by specifying s\_id in the meigen function,
as follows:

\begin{Shaded}
\begin{Highlighting}[]
\NormalTok{coords}\OtherTok{\textless{}{-}}\NormalTok{ pdat[,}\FunctionTok{c}\NormalTok{(}\StringTok{"px"}\NormalTok{, }\StringTok{"py"}\NormalTok{)]}
\NormalTok{s\_id  }\OtherTok{\textless{}{-}}\NormalTok{ pdat}\SpecialCharTok{$}\NormalTok{state}
\NormalTok{meig\_p}\OtherTok{\textless{}{-}} \FunctionTok{meigen}\NormalTok{(coords,}\AttributeTok{s\_id=}\NormalTok{s\_id)}\CommentTok{\# Moran eigenvectors by states}
\end{Highlighting}
\end{Shaded}

Currently, the following spatial panel models are available: pooling
model (no group effects); individual random effects model (state-level
group effects); time random effects model (year-level group effects);
and two-way random effects model (state and year-level group effects).
All these models consider residual spatial dependence. Here are the
codes implementing these models:

\begin{Shaded}
\begin{Highlighting}[]
\NormalTok{pmod0 }\OtherTok{\textless{}{-}} \FunctionTok{resf}\NormalTok{(}\AttributeTok{y=}\NormalTok{y,}\AttributeTok{x=}\NormalTok{x,}\AttributeTok{meig=}\NormalTok{meig\_p) }\CommentTok{\# pooling model}

\NormalTok{xgroup}\OtherTok{\textless{}{-}}\NormalTok{ pdat[,}\FunctionTok{c}\NormalTok{(}\StringTok{"state"}\NormalTok{)]}
\NormalTok{pmod1 }\OtherTok{\textless{}{-}} \FunctionTok{resf}\NormalTok{(}\AttributeTok{y=}\NormalTok{y,}\AttributeTok{x=}\NormalTok{x,}\AttributeTok{meig=}\NormalTok{meig\_p,}\AttributeTok{xgroup=}\NormalTok{xgroup)}\CommentTok{\# individual model}

\NormalTok{xgroup}\OtherTok{\textless{}{-}}\NormalTok{ pdat[,}\FunctionTok{c}\NormalTok{(}\StringTok{"year"}\NormalTok{)]}
\NormalTok{pmod2 }\OtherTok{\textless{}{-}} \FunctionTok{resf}\NormalTok{(}\AttributeTok{y=}\NormalTok{y,}\AttributeTok{x=}\NormalTok{x,}\AttributeTok{meig=}\NormalTok{meig\_p,}\AttributeTok{xgroup=}\NormalTok{xgroup)}\CommentTok{\# time model}

\NormalTok{xgroup}\OtherTok{\textless{}{-}}\NormalTok{ pdat[,}\FunctionTok{c}\NormalTok{(}\StringTok{"state"}\NormalTok{,}\StringTok{"year"}\NormalTok{)]}
\NormalTok{pmod3 }\OtherTok{\textless{}{-}} \FunctionTok{resf}\NormalTok{(}\AttributeTok{y=}\NormalTok{y,}\AttributeTok{x=}\NormalTok{x,}\AttributeTok{meig=}\NormalTok{meig\_p,}\AttributeTok{xgroup=}\NormalTok{xgroup)}\CommentTok{\# two{-}way model}
\end{Highlighting}
\end{Shaded}

Among these models, the two-way model indicates the smallest BIC. The
output is summarized as

\begin{Shaded}
\begin{Highlighting}[]
\NormalTok{pmod3}
\end{Highlighting}
\end{Shaded}

\begin{verbatim}
## Call:
## resf(y = y, x = x, xgroup = xgroup, meig = meig_p)
## 
## ----Coefficients------------------------------
##                 Estimate          SE    t_value      p_value
## (Intercept)  2.267109119 0.157658078 14.3799109 0.000000e+00
## log_pcap     0.007169765 0.023527575  0.3047388 7.606496e-01
## log_pc       0.292328541 0.022204671 13.1651825 0.000000e+00
## log_emp      0.732862585 0.024803727 29.5464701 0.000000e+00
## unemp       -0.004357505 0.001066661 -4.0851812 4.878221e-05
## 
## ----Variance parameter------------------------
## 
## Spatial effects (residuals):
##                      (Intercept)
## random_SD              0.1554446
## Moran.I/max(Moran.I)   0.3332452
## 
## Group effects:
##                state       year
## ramdom_SD 0.09486574 0.02434569
## 
## ----Error statistics--------------------------
##                      stat
## resid_SE     3.381361e-02
## adjR2(cond)  9.988953e-01
## rlogLik      1.408412e+03
## AIC         -2.796824e+03
## BIC         -2.749780e+03
## 
## NULL model: lm( y ~ x )
##    (r)loglik: 826.9817 ( AIC: -1641.963,  BIC: -1613.737 )
## 
## Note: AIC and BIC are based on the restricted/marginal likelihood.
##       Use method="ml" for comparison of models with different fixed effects (x)
\end{verbatim}

The estimated group effects are displayed as follows:

\begin{Shaded}
\begin{Highlighting}[]
\NormalTok{s\_g   }\OtherTok{\textless{}{-}}\NormalTok{ pmod3}\SpecialCharTok{$}\NormalTok{b\_g[[}\DecValTok{1}\NormalTok{]]}
\NormalTok{s\_g[}\DecValTok{1}\SpecialCharTok{:}\DecValTok{5}\NormalTok{,]}\CommentTok{\# State{-}level group effects}
\end{Highlighting}
\end{Shaded}

\begin{verbatim}
##                     Estimate         SE   t_value
## state_ALABAMA    -0.07201915 0.01388810 -5.185672
## state_ARIZONA    -0.04386270 0.01661111 -2.640564
## state_ARKANSAS   -0.07240016 0.01469587 -4.926565
## state_CALIFORNIA  0.23934902 0.01976087 12.112274
## state_COLORADO   -0.11569483 0.01232990 -9.383271
\end{verbatim}

\begin{Shaded}
\begin{Highlighting}[]
\NormalTok{t\_g   }\OtherTok{\textless{}{-}}\NormalTok{ pmod3}\SpecialCharTok{$}\NormalTok{b\_g[[}\DecValTok{2}\NormalTok{]]}
\NormalTok{t\_g[}\DecValTok{1}\SpecialCharTok{:}\DecValTok{5}\NormalTok{,]}\CommentTok{\# Year{-}level group effects}
\end{Highlighting}
\end{Shaded}

\begin{verbatim}
##               Estimate          SE    t_value
## year_1970 -0.006035536 0.011085552 -0.5444506
## year_1971  0.002885737 0.010563845  0.2731711
## year_1972  0.013268897 0.010411748  1.2744159
## year_1973  0.021939489 0.010275153  2.1351982
## year_1974 -0.009861104 0.009674704 -1.0192668
\end{verbatim}

For validation, the same panel model (but without spatial dependence) is
estimated using the plm function:

\begin{Shaded}
\begin{Highlighting}[]
\NormalTok{pm0    }\OtherTok{\textless{}{-}} \FunctionTok{plm}\NormalTok{(}\FunctionTok{log}\NormalTok{(gsp) }\SpecialCharTok{\textasciitilde{}} \FunctionTok{log}\NormalTok{(pcap) }\SpecialCharTok{+} \FunctionTok{log}\NormalTok{(pc) }\SpecialCharTok{+} \FunctionTok{log}\NormalTok{(emp) }\SpecialCharTok{+}\NormalTok{ unemp,}
              \AttributeTok{data =}\NormalTok{ pdat, }\AttributeTok{effect=}\StringTok{"twoways"}\NormalTok{,}\AttributeTok{model=}\StringTok{"random"}\NormalTok{)}
\NormalTok{pm0}
\end{Highlighting}
\end{Shaded}

\begin{verbatim}
## 
## Model Formula: log(gsp) ~ log(pcap) + log(pc) + log(emp) + unemp
## 
## Coefficients:
## (Intercept)   log(pcap)     log(pc)    log(emp)       unemp 
##   2.3634993   0.0178529   0.2655895   0.7448989  -0.0045755
\end{verbatim}

\begin{Shaded}
\begin{Highlighting}[]
\NormalTok{s\_g\_plm}\OtherTok{\textless{}{-}} \FunctionTok{ranef}\NormalTok{(pm0,}\StringTok{"individual"}\NormalTok{)}
\NormalTok{t\_g\_plm}\OtherTok{\textless{}{-}} \FunctionTok{ranef}\NormalTok{(pm0,}\StringTok{"time"}\NormalTok{)}
\end{Highlighting}
\end{Shaded}

The coefficient estimates are similar. The plots below compare estimated
group effects. Estimated state-level effects have differences because
our models consider residual spatial dependence, while plm does not (by
default). Time effects are quite similar.

\begin{Shaded}
\begin{Highlighting}[]
\FunctionTok{plot}\NormalTok{(s\_g\_plm,s\_g[,}\DecValTok{1}\NormalTok{],}\AttributeTok{xlab=}\StringTok{"plm"}\NormalTok{,}\AttributeTok{ylab=}\StringTok{"resf"}\NormalTok{)}
\FunctionTok{abline}\NormalTok{(}\DecValTok{0}\NormalTok{,}\DecValTok{1}\NormalTok{,}\AttributeTok{col=}\StringTok{"red"}\NormalTok{)}
\end{Highlighting}
\end{Shaded}

\includegraphics{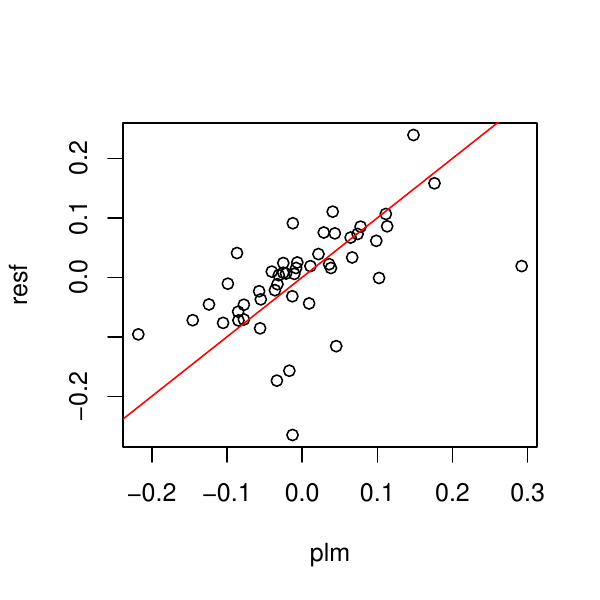}

\begin{Shaded}
\begin{Highlighting}[]
\FunctionTok{plot}\NormalTok{(t\_g\_plm,t\_g[,}\DecValTok{1}\NormalTok{],}\AttributeTok{xlab=}\StringTok{"plm"}\NormalTok{,}\AttributeTok{ylab=}\StringTok{"resf"}\NormalTok{)}
\FunctionTok{abline}\NormalTok{(}\DecValTok{0}\NormalTok{,}\DecValTok{1}\NormalTok{,}\AttributeTok{col=}\StringTok{"red"}\NormalTok{)}
\end{Highlighting}
\end{Shaded}

\includegraphics{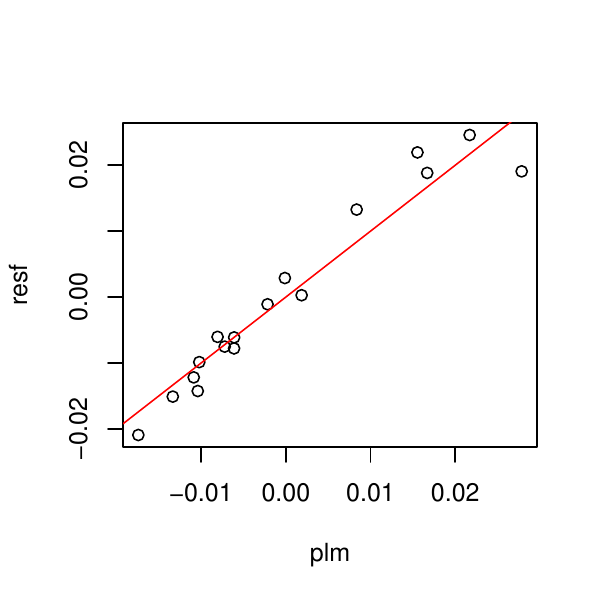}

\hypertarget{spatial-prediction}{%
\subsection{Spatial prediction}\label{spatial-prediction}}

This package provides functions for ESF/RE-ESF-based spatial
interpolation minimizing the expected prediction error (just like
kriging). RE-ESF approximates a Gaussian process or the kriging model,
which has actively been used for spatial prediction, and ESF is a
special case (Murakami and Griffith, 2015). Because ESF and RE-ESF
impose approximations, they are faster for very large samples.

In this tutorial, the Lucas housing price data with sample size being
25,357 is used. In the prediction, ``price'' is used as the explained
variable, and ``age,'' ``rooms,'' ``beds,'' and ``year'' are used as
covariates.

\begin{Shaded}
\begin{Highlighting}[]
\FunctionTok{require}\NormalTok{(spData)}
\FunctionTok{data}\NormalTok{(house)}
\NormalTok{dat0  }\OtherTok{\textless{}{-}} \FunctionTok{st\_as\_sf}\NormalTok{(house)}
\NormalTok{dat   }\OtherTok{\textless{}{-}} \FunctionTok{data.frame}\NormalTok{(}\FunctionTok{st\_coordinates}\NormalTok{(dat0), dat0[,}\FunctionTok{c}\NormalTok{(}\StringTok{"price"}\NormalTok{,}\StringTok{"age"}\NormalTok{,}\StringTok{"rooms"}\NormalTok{,}\StringTok{"beds"}\NormalTok{,}\StringTok{"syear"}\NormalTok{)])}
\end{Highlighting}
\end{Shaded}

A total of 20,000 randomly selected samples are used for model
estimation, and the other 5,357 samples are used for accuracy
evaluation. The code below creates the data for observation sites
(coords, y, x) and for unobserved sites (coords0, y0, x0):

\begin{Shaded}
\begin{Highlighting}[]
\NormalTok{samp    }\OtherTok{\textless{}{-}} \FunctionTok{sample}\NormalTok{(}\FunctionTok{dim}\NormalTok{(dat)[}\DecValTok{1}\NormalTok{], }\DecValTok{20000}\NormalTok{)}
\NormalTok{coords}\OtherTok{\textless{}{-}}\NormalTok{ dat[samp ,}\FunctionTok{c}\NormalTok{(}\StringTok{"X"}\NormalTok{,}\StringTok{"Y"}\NormalTok{)]}
\NormalTok{y       }\OtherTok{\textless{}{-}} \FunctionTok{log}\NormalTok{(dat[samp,}\StringTok{"price"}\NormalTok{])}
\NormalTok{x     }\OtherTok{\textless{}{-}}\NormalTok{ dat[samp,}\FunctionTok{c}\NormalTok{(}\StringTok{"age"}\NormalTok{,}\StringTok{"rooms"}\NormalTok{,}\StringTok{"beds"}\NormalTok{,}\StringTok{"syear"}\NormalTok{)]}

\NormalTok{coords0}\OtherTok{\textless{}{-}}\NormalTok{ dat[}\SpecialCharTok{{-}}\NormalTok{samp ,}\FunctionTok{c}\NormalTok{(}\StringTok{"X"}\NormalTok{,}\StringTok{"Y"}\NormalTok{)]}
\NormalTok{y0    }\OtherTok{\textless{}{-}} \FunctionTok{log}\NormalTok{(dat[}\SpecialCharTok{{-}}\NormalTok{samp,}\StringTok{"price"}\NormalTok{]) }\CommentTok{\# for valudation}
\NormalTok{x0    }\OtherTok{\textless{}{-}}\NormalTok{ dat[}\SpecialCharTok{{-}}\NormalTok{samp,}\FunctionTok{c}\NormalTok{(}\StringTok{"age"}\NormalTok{,}\StringTok{"rooms"}\NormalTok{,}\StringTok{"beds"}\NormalTok{,}\StringTok{"syear"}\NormalTok{)]}
\end{Highlighting}
\end{Shaded}

The prediction is done in two steps: (1) evaluation of Moran
eigenvectors at prediction sites using the meigen0 function; (2)
prediction using the predict0 function. Below is a sample code based on
the resf function:

\begin{Shaded}
\begin{Highlighting}[]
\NormalTok{start.time1}\OtherTok{\textless{}{-}}\FunctionTok{proc.time}\NormalTok{()}\DocumentationTok{\#\#\#\#\#\# For CP time evaluation}
\NormalTok{meig     }\OtherTok{\textless{}{-}} \FunctionTok{meigen\_f}\NormalTok{(coords)}
\NormalTok{meig0    }\OtherTok{\textless{}{-}} \FunctionTok{meigen0}\NormalTok{( }\AttributeTok{meig=}\NormalTok{meig, }\AttributeTok{coords0=}\NormalTok{coords0 )}
\NormalTok{mod      }\OtherTok{\textless{}{-}} \FunctionTok{resf}\NormalTok{( }\AttributeTok{y =}\NormalTok{ y, }\AttributeTok{x =}\NormalTok{ x, }\AttributeTok{meig =}\NormalTok{ meig )}
\NormalTok{pred0      }\OtherTok{\textless{}{-}} \FunctionTok{predict0}\NormalTok{( }\AttributeTok{mod =}\NormalTok{ mod, }\AttributeTok{x0 =}\NormalTok{ x0, }\AttributeTok{meig0=}\NormalTok{meig0 )}
\NormalTok{end.time1}\OtherTok{\textless{}{-}} \FunctionTok{proc.time}\NormalTok{()}\DocumentationTok{\#\#\#\#\#\# For CP time evaluation}
\end{Highlighting}
\end{Shaded}

Note that the first and last lines are just for computing time
evaluation. NVCs are considered if adding NVC=TRUE in the resf function.
The meigen\_f function is used for fast computation.

The outputs shown below include predicted values (pred), predicted trend
(xb), and predicted residual spatial component (sf\_residuals).

\begin{Shaded}
\begin{Highlighting}[]
\NormalTok{pred0}\SpecialCharTok{$}\NormalTok{pred[}\DecValTok{1}\SpecialCharTok{:}\DecValTok{5}\NormalTok{,]}
\end{Highlighting}
\end{Shaded}

\begin{verbatim}
##        pred       xb sf_residual
## 5  11.81208 11.34347   0.4686106
## 11 11.62800 11.19501   0.4329969
## 15 11.38538 10.89625   0.4891365
## 16 11.77415 11.36873   0.4054135
## 21 11.08661 10.67609   0.4105252
\end{verbatim}

\begin{Shaded}
\begin{Highlighting}[]
\NormalTok{pred     }\OtherTok{\textless{}{-}}\NormalTok{ pred0}\SpecialCharTok{$}\NormalTok{pred[,}\DecValTok{1}\NormalTok{]}
\end{Highlighting}
\end{Shaded}

On the other hand, here is a code for a spatial prediction based on an
S(N)VC model:

\begin{Shaded}
\begin{Highlighting}[]
\NormalTok{start.time2}\OtherTok{\textless{}{-}}\FunctionTok{proc.time}\NormalTok{()}\DocumentationTok{\#\#\#\#\#\# For CP time evaluation}
\NormalTok{meig     }\OtherTok{\textless{}{-}} \FunctionTok{meigen\_f}\NormalTok{(coords)}
\NormalTok{meig0    }\OtherTok{\textless{}{-}} \FunctionTok{meigen0}\NormalTok{( }\AttributeTok{meig=}\NormalTok{meig, }\AttributeTok{coords0=}\NormalTok{coords0 )}
\NormalTok{mod2       }\OtherTok{\textless{}{-}} \FunctionTok{resf\_vc}\NormalTok{( }\AttributeTok{y =}\NormalTok{ y, }\AttributeTok{x =}\NormalTok{ x, }\AttributeTok{meig =}\NormalTok{ meig )}
\end{Highlighting}
\end{Shaded}

\begin{verbatim}
## [1] "-------  Iteration 1  -------"
## [1] "1/5"
## [1] "2/5"
## [1] "3/5"
## [1] "4/5"
## [1] "5/5"
## [1] "BIC: 13501.483"
## [1] "-------  Iteration 2  -------"
## [1] "1/5"
## [1] "2/5"
## [1] "3/5"
## [1] "4/5"
## [1] "5/5"
## [1] "BIC: 13147.54"
## [1] "-------  Iteration 3  -------"
## [1] "1/5"
## [1] "2/5"
## [1] "3/5"
## [1] "4/5"
## [1] "5/5"
## [1] "BIC: 13144.252"
## [1] "-------  Iteration 4  -------"
## [1] "1/5"
## [1] "2/5"
## [1] "3/5"
## [1] "4/5"
## [1] "5/5"
## [1] "BIC: 13144.219"
## [1] "-------  Iteration 5  -------"
## [1] "1/5"
## [1] "2/5"
## [1] "3/5"
## [1] "4/5"
## [1] "5/5"
## [1] "BIC: 13144.219"
## [1] "-------  Iteration 6  -------"
## [1] "1/5"
## [1] "2/5"
## [1] "3/5"
## [1] "4/5"
## [1] "5/5"
## [1] "BIC: 13144.219"
\end{verbatim}

\begin{Shaded}
\begin{Highlighting}[]
\NormalTok{pred02   }\OtherTok{\textless{}{-}} \FunctionTok{predict0}\NormalTok{( }\AttributeTok{mod =}\NormalTok{ mod2, }\AttributeTok{x0 =}\NormalTok{ x0, }\AttributeTok{meig0=}\NormalTok{meig0 )}
\NormalTok{end.time2}\OtherTok{\textless{}{-}} \FunctionTok{proc.time}\NormalTok{()}\DocumentationTok{\#\#\#\#\#\# For CP time evaluation}
\end{Highlighting}
\end{Shaded}

NVCs are considered by adding NVC=TRUE in the resf\_vc function. Here
are the output variables:

\begin{Shaded}
\begin{Highlighting}[]
\NormalTok{pred02}\SpecialCharTok{$}\NormalTok{pred[}\DecValTok{1}\SpecialCharTok{:}\DecValTok{5}\NormalTok{,]}
\end{Highlighting}
\end{Shaded}

\begin{verbatim}
##        pred       xb sf_residual
## 5  11.68471 11.64464  0.04006557
## 11 11.57193 11.54166  0.03027240
## 15 11.41369 11.36237  0.05131992
## 16 11.61277 11.57533  0.03744066
## 21 11.13501 11.09377  0.04124334
\end{verbatim}

\begin{Shaded}
\begin{Highlighting}[]
\NormalTok{pred2    }\OtherTok{\textless{}{-}}\NormalTok{ pred02}\SpecialCharTok{$}\NormalTok{pred[,}\DecValTok{1}\NormalTok{]}
\end{Highlighting}
\end{Shaded}

The root mean squared prediction error (RMSPE) and the computational
time of the spatial regression model (resf) are as follows:

\begin{Shaded}
\begin{Highlighting}[]
\FunctionTok{sqrt}\NormalTok{(}\FunctionTok{sum}\NormalTok{((pred}\SpecialCharTok{{-}}\NormalTok{y0)}\SpecialCharTok{\^{}}\DecValTok{2}\NormalTok{)}\SpecialCharTok{/}\FunctionTok{length}\NormalTok{(y0))}\CommentTok{\#rmse}
\end{Highlighting}
\end{Shaded}

\begin{verbatim}
## [1] 0.3509916
\end{verbatim}

\begin{Shaded}
\begin{Highlighting}[]
\NormalTok{(end.time1 }\SpecialCharTok{{-}}\NormalTok{ start.time1)[}\DecValTok{3}\NormalTok{]}\CommentTok{\#computational time (second)}
\end{Highlighting}
\end{Shaded}

\begin{verbatim}
## elapsed 
##     8.8
\end{verbatim}

while those of the SVC model (resf\_vc) are as follows:

\begin{Shaded}
\begin{Highlighting}[]
\FunctionTok{sqrt}\NormalTok{(}\FunctionTok{sum}\NormalTok{((pred2}\SpecialCharTok{{-}}\NormalTok{y0)}\SpecialCharTok{\^{}}\DecValTok{2}\NormalTok{)}\SpecialCharTok{/}\FunctionTok{length}\NormalTok{(y0))}\CommentTok{\#rmse}
\end{Highlighting}
\end{Shaded}

\begin{verbatim}
## [1] 0.3385873
\end{verbatim}

\begin{Shaded}
\begin{Highlighting}[]
\NormalTok{(end.time2 }\SpecialCharTok{{-}}\NormalTok{ start.time2)[}\DecValTok{3}\NormalTok{]}\CommentTok{\#computational time (second)}
\end{Highlighting}
\end{Shaded}

\begin{verbatim}
## elapsed 
##  131.33
\end{verbatim}

The results suggest that both models are available for large samples. It
is also demonstrated that while the spatial regression model is faster
than the SVC model, the SVC model is slightly more accurate. The actual
values (y0) and predicted values (pred/pred2) are compared below:

\begin{Shaded}
\begin{Highlighting}[]
\FunctionTok{par}\NormalTok{(}\AttributeTok{mfrow=}\FunctionTok{c}\NormalTok{(}\DecValTok{1}\NormalTok{,}\DecValTok{2}\NormalTok{)) }
\FunctionTok{plot}\NormalTok{(y0,pred);}\FunctionTok{abline}\NormalTok{(}\DecValTok{0}\NormalTok{,}\DecValTok{1}\NormalTok{,}\AttributeTok{col=}\StringTok{"red"}\NormalTok{)}
\FunctionTok{plot}\NormalTok{(y0,pred2);}\FunctionTok{abline}\NormalTok{(}\DecValTok{0}\NormalTok{,}\DecValTok{1}\NormalTok{,}\AttributeTok{col=}\StringTok{"red"}\NormalTok{)}
\end{Highlighting}
\end{Shaded}

\includegraphics{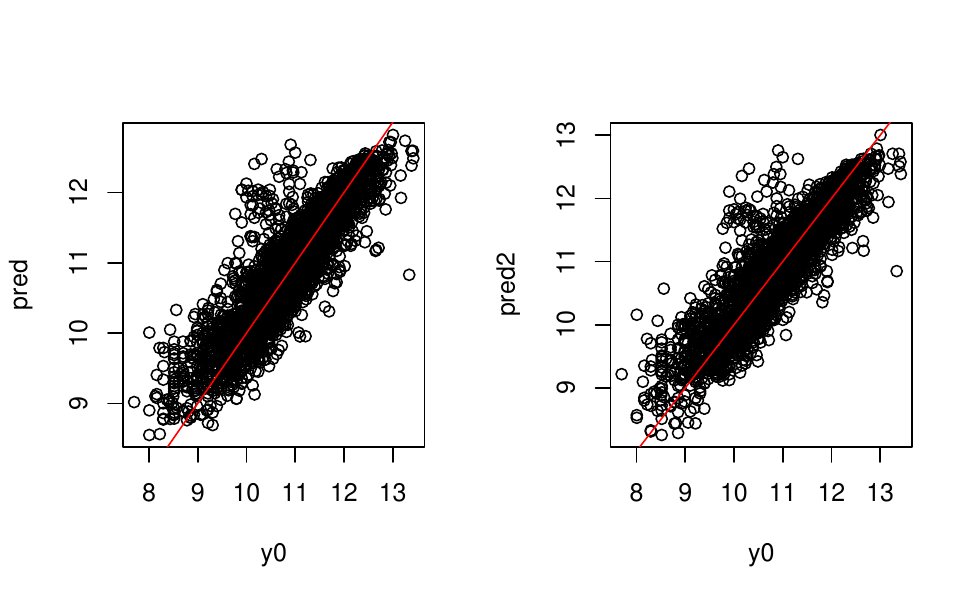}

(RE-)ESF considers a limited number of eigenvectors, which limits the
model flexibility. Because of that, (RE-)ESF suffers from a
degeneracy/over-smoothing problem that decreases modeling accuracy for
large samples. The addlearn\_local function is useful to address this
problem. This function estimates an improved SVC model by
aggregating/averaging the pre-estimated SVC model (i.e., mod2) with
local SVC models that are estimated by k-means-based spatial clusters
each of which contains roughly 600 samples (see Murakami et al., 2023).
Unlike the resf\_vc and/or besf\_vc function, the improved SVC model
considers not only (global) eigenvectors but also local eigenvectors;
the resulting model accurately captures local patterns even from very
large samples.

Here is a sample code for the model aggregation and prediction after the
model aggregation:

\begin{Shaded}
\begin{Highlighting}[]
\NormalTok{start.time3}\OtherTok{\textless{}{-}}\FunctionTok{proc.time}\NormalTok{()  }\DocumentationTok{\#\#\#\#\#\# For CP time evaluation}
\NormalTok{mod3       }\OtherTok{\textless{}{-}} \FunctionTok{addlearn\_local}\NormalTok{(}\AttributeTok{mod=}\NormalTok{mod2, }\AttributeTok{meig0 =}\NormalTok{ meig0, }\AttributeTok{x0 =}\NormalTok{ x0)}
\end{Highlighting}
\end{Shaded}

\begin{verbatim}
## [1] "-------- Aggregating 36 local sub-models ---------"
## [1] "1/36"
## [1] "2/36"
## [1] "3/36"
## [1] "4/36"
## [1] "5/36"
## [1] "6/36"
## [1] "7/36"
## [1] "8/36"
## [1] "9/36"
## [1] "10/36"
## [1] "11/36"
## [1] "12/36"
## [1] "13/36"
## [1] "14/36"
## [1] "15/36"
## [1] "16/36"
## [1] "17/36"
## [1] "18/36"
## [1] "19/36"
## [1] "20/36"
## [1] "21/36"
## [1] "22/36"
## [1] "23/36"
## [1] "24/36"
## [1] "25/36"
## [1] "26/36"
## [1] "27/36"
## [1] "28/36"
## [1] "29/36"
## [1] "30/36"
## [1] "31/36"
## [1] "32/36"
## [1] "33/36"
## [1] "34/36"
## [1] "35/36"
## [1] "36/36"
\end{verbatim}

\begin{Shaded}
\begin{Highlighting}[]
\NormalTok{pred3      }\OtherTok{\textless{}{-}}\NormalTok{ mod3}\SpecialCharTok{$}\NormalTok{pred0[,}\DecValTok{1}\NormalTok{]}
\NormalTok{end.time3  }\OtherTok{\textless{}{-}} \FunctionTok{proc.time}\NormalTok{() }\DocumentationTok{\#\#\#\#\#\# For CP time evaluation}
\end{Highlighting}
\end{Shaded}

The resulting RMSE is confirmed to be smaller than mod2, which is before
the model aggregation:

\begin{Shaded}
\begin{Highlighting}[]
\FunctionTok{sqrt}\NormalTok{(}\FunctionTok{sum}\NormalTok{((pred3}\SpecialCharTok{{-}}\NormalTok{y0)}\SpecialCharTok{\^{}}\DecValTok{2}\NormalTok{)}\SpecialCharTok{/}\FunctionTok{length}\NormalTok{(y0))}\CommentTok{\#rmse}
\end{Highlighting}
\end{Shaded}

\begin{verbatim}
## [1] 0.3128213
\end{verbatim}

\begin{Shaded}
\begin{Highlighting}[]
\NormalTok{(end.time3 }\SpecialCharTok{{-}}\NormalTok{ start.time3)[}\DecValTok{3}\NormalTok{]}\CommentTok{\#computational time (second)}
\end{Highlighting}
\end{Shaded}

\begin{verbatim}
## elapsed 
##  215.62
\end{verbatim}

While the addlearn\_local function requires an additional computation
time, it can be paralleled by specifying parallel = TRUE. The accuracy
difference between the models with/without the model
aggregation/averaging increases as the sample size increases. This
addlearn\_local function is especially recommended for larger samples.

The addlearn\_local function is also useful to improve SVC coefficient
estimation accuracy as demonstrated in Murakami et al.~(2023). See
Section 6.3 for further detail.

\hypertarget{non-gaussian-spatial-regression-models}{%
\section{Non-Gaussian spatial regression
models}\label{non-gaussian-spatial-regression-models}}

This package is now available for modeling a wide variety of
non-Gaussian data including count data. Unlike the conventional
generalized linear model (GLM), the implemented model estimates the most
likely data distribution (i.e., probability density/mass function)
without explicitly specifying the data distribution (see Murakami et
al., 2021). See Murakami (2021) or vignette\_spmoran(nongaussian).pdf,
which is another vignette in the same GitHub page
\url{https://github.com/dmuraka/spmoran} for details on how to implement
it.

\hypertarget{spatially-filtered-unconditional-quantile-regression}{%
\section{Spatially filtered unconditional quantile
regression}\label{spatially-filtered-unconditional-quantile-regression}}

While the usual (conditional) quantile regression (CQR) estimates the
influence of \(x_k\) on the \(\tau\)-th conditional quantile of \(y\),
\(q_\tau(y|x_k)\), the unconditional quantile regression estimates the
influence of \(x_k\) on the ``unconditional'' quantile of y,
\(q_\tau(y)\) (Firpo et al., 2009).

Suppose that \(y\) and \(x_k\) represent land price and accessibility,
respectively. UQR estimates the influence of accessibility on land price
by quantile; it is interpretable and useful for hedonic land price
analysis, for example. By contrast, this interpretation does not hold
for CQR because it estimates the influence of accessibility on
conditional land prices (land price conditional on explanatory
variables). Higher conditional land price does not mean higher land
price; rather, it means overprice relative to the price expected by the
explanatory variables. Therefore, CQR has difficulty in its
interpretation, in some cases, including hedonic land price modeling.

The spatial filter UQR (SF-UQR) model (Murakami and Seya, 2019), which
is implemented in this package, is formulated as
\[q_\tau(y_i)=\sum^K_{k=1}x_{i,k}\beta_{k,\tau}+f_{MC,\tau}(s_i)+\epsilon_{i,\tau}, \hspace{0.5cm}\epsilon_{i,\tau} \sim N(0, \sigma^2_\tau),\]
This model is a UQR considering spatial dependence.

The resf\_qr function implements this model. Below is a sample code. If
boot=TRUE in resf\_qr, a semiparametric bootstrapping is performed to
estimate the standard errors of the regression coefficients. By default,
this function estimates models at 0.1, 0.2,\ldots, 0.9 quantiles.

\begin{Shaded}
\begin{Highlighting}[]
\NormalTok{y       }\OtherTok{\textless{}{-}}\NormalTok{ boston.c[, }\StringTok{"CMEDV"}\NormalTok{ ]}
\NormalTok{x       }\OtherTok{\textless{}{-}}\NormalTok{ boston.c[,}\FunctionTok{c}\NormalTok{(}\StringTok{"CRIM"}\NormalTok{,}\StringTok{"ZN"}\NormalTok{,}\StringTok{"INDUS"}\NormalTok{, }\StringTok{"CHAS"}\NormalTok{, }\StringTok{"NOX"}\NormalTok{,}\StringTok{"RM"}\NormalTok{, }\StringTok{"AGE"}\NormalTok{)]}
\NormalTok{coords}\OtherTok{\textless{}{-}}\NormalTok{ boston.c[,}\FunctionTok{c}\NormalTok{(}\StringTok{"LON"}\NormalTok{,}\StringTok{"LAT"}\NormalTok{)]}
\NormalTok{meig    }\OtherTok{\textless{}{-}} \FunctionTok{meigen}\NormalTok{(}\AttributeTok{coords=}\NormalTok{coords)}
\NormalTok{res   }\OtherTok{\textless{}{-}} \FunctionTok{resf\_qr}\NormalTok{(}\AttributeTok{y=}\NormalTok{y,}\AttributeTok{x=}\NormalTok{x,}\AttributeTok{meig=}\NormalTok{meig, }\AttributeTok{boot=}\ConstantTok{TRUE}\NormalTok{)}
\end{Highlighting}
\end{Shaded}

\begin{verbatim}
## [1] "------- Complete: tau=0.1 -------"
## [1] "------- Complete: tau=0.2 -------"
## [1] "------- Complete: tau=0.3 -------"
## [1] "------- Complete: tau=0.4 -------"
## [1] "------- Complete: tau=0.5 -------"
## [1] "------- Complete: tau=0.6 -------"
## [1] "------- Complete: tau=0.7 -------"
## [1] "------- Complete: tau=0.8 -------"
## [1] "------- Complete: tau=0.9 -------"
\end{verbatim}

Here is a summary of the estimation result:

\begin{Shaded}
\begin{Highlighting}[]
\NormalTok{res}
\end{Highlighting}
\end{Shaded}

\begin{verbatim}
## Call:
## resf_qr(y = y, x = x, meig = meig, boot = TRUE)
## 
## ----Coefficients------------------------------
##                  tau=0.1      tau=0.2       tau=0.3      tau=0.4       tau=0.5
## (Intercept)  23.86841970  29.16185736  26.550125353  21.16263694  17.151053980
## CRIM         -0.36845124  -0.21172051  -0.106949379  -0.08357496  -0.070290258
## ZN           -0.01169653  -0.01627637  -0.009652286  -0.01947512  -0.008198579
## INDUS         0.25009373   0.03992002  -0.111010420  -0.01521113  -0.096468769
## CHAS          0.98647836   1.28770409   0.438428954   0.26777796  -0.048278485
## NOX         -32.89857783 -23.60303480 -15.109338348 -12.70090129 -11.263158727
## RM            0.71728433   0.49201634   1.169115918   2.21382993   3.004059676
## AGE           0.01977978  -0.05087471  -0.082548477  -0.11192561  -0.105681036
##                   tau=0.6      tau=0.7     tau=0.8      tau=0.9
## (Intercept)  13.999671526  11.28433168 -23.3939330 -57.24239068
## CRIM         -0.064412593  -0.07823561  -0.1876252  -0.18934294
## ZN            0.007962903   0.01009742   0.1635369   0.03890142
## INDUS        -0.167039581  -0.30344029  -0.9074079  -0.49797629
## CHAS         -1.665298913  -1.51518801  -3.8773852  -0.04635798
## NOX         -11.405913169 -20.36309658 -39.1980207 -41.26421537
## RM            3.730680883   5.25253569  13.7698457  19.62200618
## AGE          -0.092068861  -0.07567382  -0.0587608  -0.03904752
## 
## ----Spatial effects (residuals)---------------
##                               tau=0.1   tau=0.2   tau=0.3   tau=0.4   tau=0.5
## spcomp_SD                   7.1522586 8.1254770 5.7952363 4.4135132 4.7198329
## spcomp_Moran.I/max(Moran.I) 0.2375865 0.3228553 0.3239407 0.3650454 0.5096847
##                               tau=0.6   tau=0.7    tau=0.8    tau=0.9
## spcomp_SD                   4.8818059 6.3633073 16.9989855 16.3826940
## spcomp_Moran.I/max(Moran.I) 0.5690447 0.6935049  0.6757823  0.7203891
## 
## ----Error statistics--------------------------
##                     tau=0.1   tau=0.2  tau=0.3   tau=0.4   tau=0.5   tau=0.6
## resid_SE          6.4395412 6.2086846 5.169030 4.7999618 4.5977255 4.8160068
## quasi_adjR2(cond) 0.6007294 0.6828421 0.666506 0.6183801 0.6229795 0.6121279
##                     tau=0.7    tau=0.8    tau=0.9
## resid_SE          5.6288391 12.2961444 18.6716254
## quasi_adjR2(cond) 0.6153019  0.6741455  0.4582676
\end{verbatim}

The estimated coefficients can be visualized using the plot\_qr
function, as below. The numbers 1 to 5 specify which coefficients are
plotted (1: intercept). In each panel, solid lines are estimated
coefficients, and gray areas are their 95\% confidence intervals.

\begin{Shaded}
\begin{Highlighting}[]
\FunctionTok{plot\_qr}\NormalTok{( res, }\DecValTok{1}\NormalTok{ )}
\end{Highlighting}
\end{Shaded}

\includegraphics{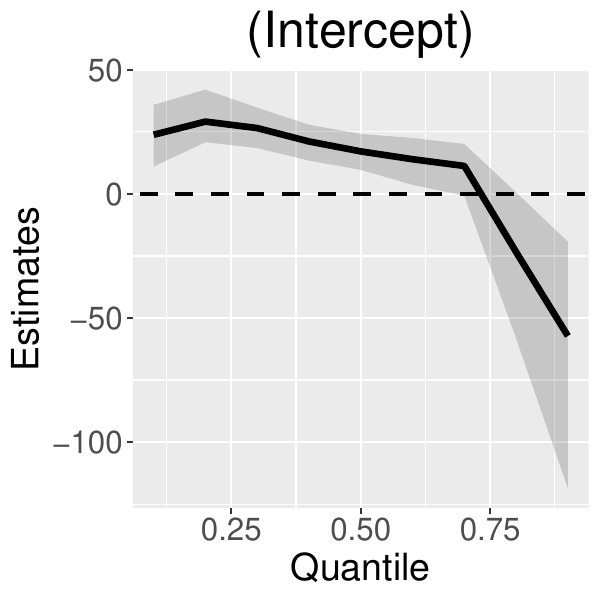}

\begin{Shaded}
\begin{Highlighting}[]
\FunctionTok{plot\_qr}\NormalTok{( res, }\DecValTok{2}\NormalTok{ )}
\end{Highlighting}
\end{Shaded}

\includegraphics{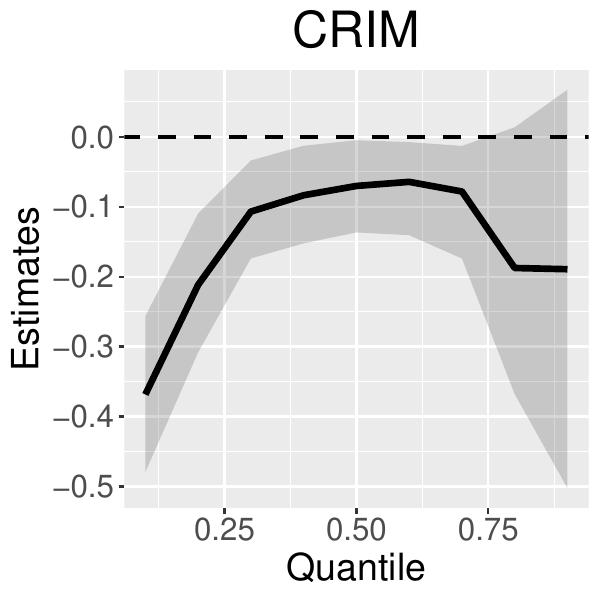}

\begin{Shaded}
\begin{Highlighting}[]
\FunctionTok{plot\_qr}\NormalTok{( res, }\DecValTok{3}\NormalTok{ )}
\end{Highlighting}
\end{Shaded}

\includegraphics{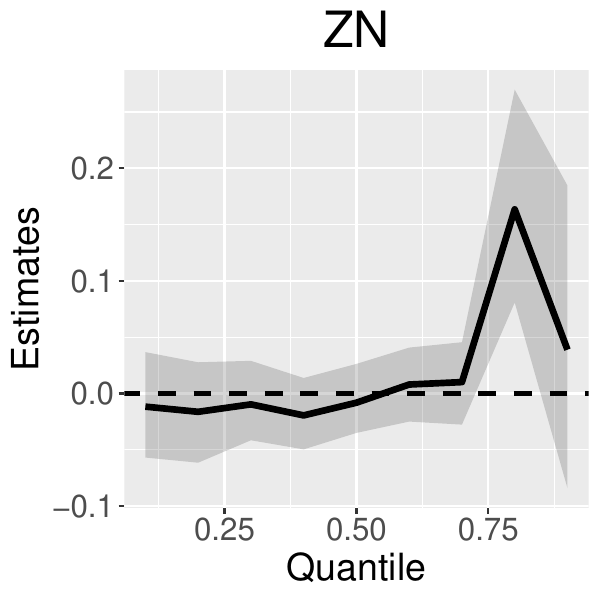}

\begin{Shaded}
\begin{Highlighting}[]
\FunctionTok{plot\_qr}\NormalTok{( res, }\DecValTok{4}\NormalTok{ )}
\end{Highlighting}
\end{Shaded}

\includegraphics{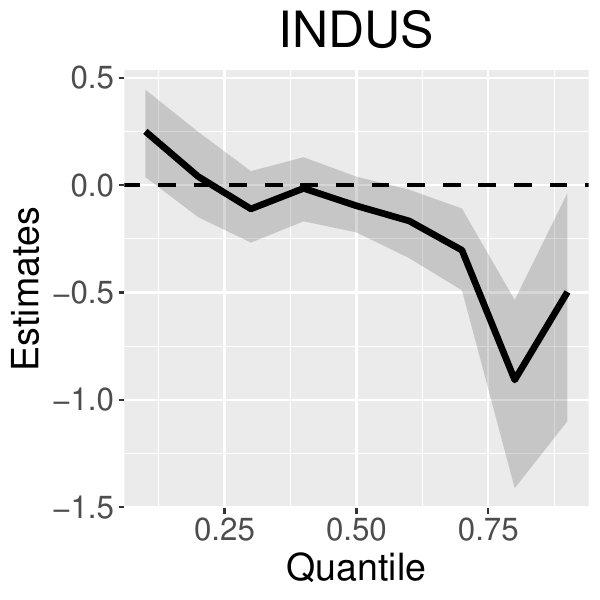}

\begin{Shaded}
\begin{Highlighting}[]
\FunctionTok{plot\_qr}\NormalTok{( res, }\DecValTok{5}\NormalTok{ )}
\end{Highlighting}
\end{Shaded}

\includegraphics{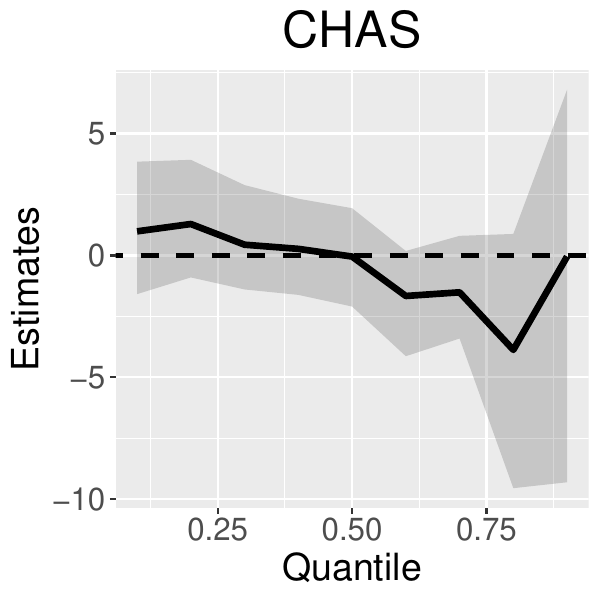}

Standard errors and the scaled Moran coefficient (Moran.I/max(Moran.I)),
which is a measure of spatial scale by quantile, are plotted if par =
``s'' is added. Here are the plots:

\begin{Shaded}
\begin{Highlighting}[]
\FunctionTok{plot\_qr}\NormalTok{( res, }\AttributeTok{par =} \StringTok{"s"}\NormalTok{ , }\DecValTok{1}\NormalTok{ )}
\end{Highlighting}
\end{Shaded}

\includegraphics{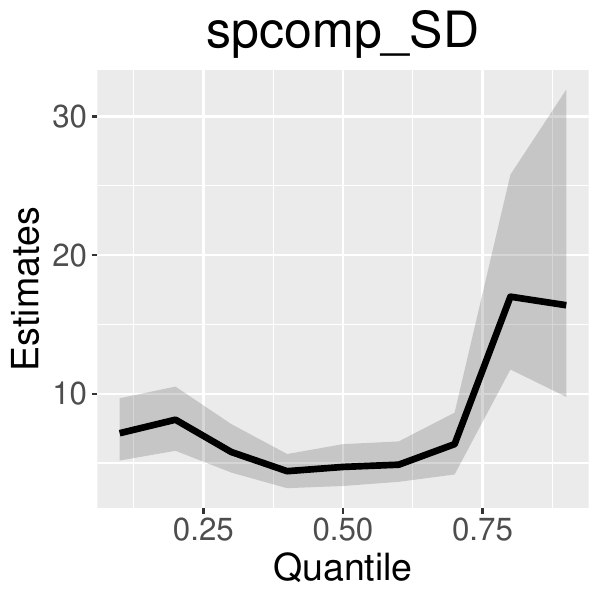}

\begin{Shaded}
\begin{Highlighting}[]
\FunctionTok{plot\_qr}\NormalTok{( res, }\AttributeTok{par =} \StringTok{"s"}\NormalTok{ , }\DecValTok{2}\NormalTok{ )}
\end{Highlighting}
\end{Shaded}

\includegraphics{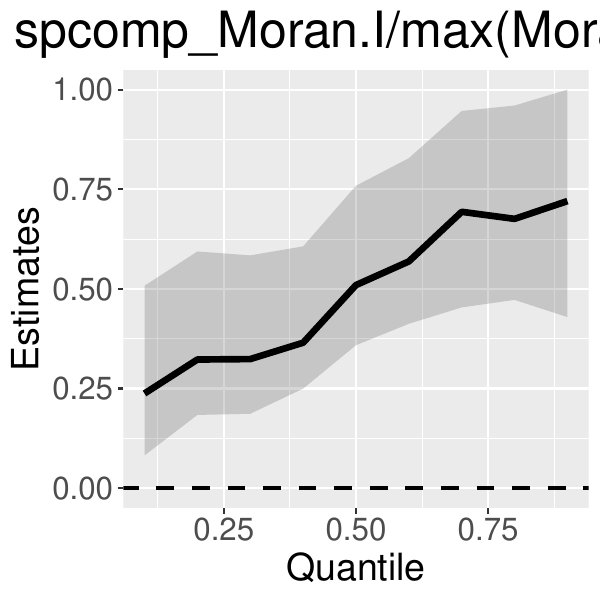}

\hypertarget{low-rank-spatial-econometric-models}{%
\section{Low rank spatial econometric
models}\label{low-rank-spatial-econometric-models}}

While ESF/RE-ESF and their extensions approximate Gaussian processes,
this section explains low rank spatial econometric models approximating
spatial econometric models (see Murakami et al., 2018).

\hypertarget{spatial-weight-matrix-and-their-eigenvectors}{%
\subsection{Spatial weight matrix and their
eigenvectors}\label{spatial-weight-matrix-and-their-eigenvectors}}

The low rank models use eigenvectors and eigenvalues of a spatial
connectivity matrix, which is called a spatial weight matrix or W matrix
in spatial econometrics. The weigen function is available for the
eigen-decomposition. Here is a code extracting the eigenvectors and
eigenvalues from spatial polygons:

\begin{Shaded}
\begin{Highlighting}[]
\FunctionTok{data}\NormalTok{( boston )}
\NormalTok{poly    }\OtherTok{\textless{}{-}} \FunctionTok{st\_read}\NormalTok{( }\FunctionTok{system.file}\NormalTok{( }\StringTok{"shapes/boston\_tracts.shp"}\NormalTok{, }\AttributeTok{package =} \StringTok{"spData"}\NormalTok{ )[ }\DecValTok{1}\NormalTok{ ] )}
\end{Highlighting}
\end{Shaded}

\begin{verbatim}
## Reading layer `boston_tracts' from data source 
##   `C:\Users\dmura\AppData\Local\R\win-library\4.3\spData\shapes\boston_tracts.shp' 
##   using driver `ESRI Shapefile'
## Simple feature collection with 506 features and 36 fields
## Geometry type: POLYGON
## Dimension:     XY
## Bounding box:  xmin: -71.52311 ymin: 42.00305 xmax: -70.63823 ymax: 42.67307
## Geodetic CRS:  NAD27
\end{verbatim}

\begin{Shaded}
\begin{Highlighting}[]
\NormalTok{weig    }\OtherTok{\textless{}{-}} \FunctionTok{weigen}\NormalTok{( poly )           }\DocumentationTok{\#\#\#\# Rook adjacency{-}based W}
\end{Highlighting}
\end{Shaded}

By default, the weigen function returns a Rook adjacency-based W matrix.
Other than that, knn-based W, Delaunay triangulation-based W, and
user-specified W are also available.

\hypertarget{models}{%
\subsection{Models}\label{models}}

\hypertarget{low-rank-spatial-lag-model}{%
\subsubsection{Low rank spatial lag
model}\label{low-rank-spatial-lag-model}}

The low rank spatial lag model (LSLM) approximates the following model:
\[y_i=\beta_0+z_i+\epsilon_i\hspace{0.5cm}\epsilon_i \sim N(0,\sigma^2)\\
z_i=\rho\sum^N_{i \neq j}w_{i,j}z_j+\sum^K_{k \neq 1}x_{i,k}\beta_k+u_i \hspace{0.5cm} u_i \sim N(0,\tau^2)\]
where \(z_i\) is defined by the classical spatial lag model (SLM; see
LeSage and Pace, 2009) with parameters \(\rho\) and \(\tau^2\). Just
like the original SLM, \(\rho\) takes a value between 1 and
\(1/\lambda_N (< 0)\). Larger positive \(\rho\) means stronger positive
dependence. \(\tau^2\) represents the variance of the SLM-based spatial
process (i.e., \(z_i\)), while \(\sigma^2\) represents the variance of
the data noise \(\epsilon_i\). Because of the additional noise term, the
LSLM estimates are different from the original SLM, in particular if
data is noisy.

The LSLM is implemented using the lslm function. Here is a sample code:

\begin{Shaded}
\begin{Highlighting}[]
\NormalTok{y       }\OtherTok{\textless{}{-}}\NormalTok{ boston.c[, }\StringTok{"CMEDV"}\NormalTok{ ]}
\NormalTok{x       }\OtherTok{\textless{}{-}}\NormalTok{ boston.c[,}\FunctionTok{c}\NormalTok{(}\StringTok{"CRIM"}\NormalTok{,}\StringTok{"ZN"}\NormalTok{,}\StringTok{"INDUS"}\NormalTok{, }\StringTok{"CHAS"}\NormalTok{, }\StringTok{"NOX"}\NormalTok{,}\StringTok{"RM"}\NormalTok{, }\StringTok{"AGE"}\NormalTok{)]}
\NormalTok{coords}\OtherTok{\textless{}{-}}\NormalTok{ boston.c[,}\FunctionTok{c}\NormalTok{(}\StringTok{"LON"}\NormalTok{,}\StringTok{"LAT"}\NormalTok{)]}
\NormalTok{res   }\OtherTok{\textless{}{-}} \FunctionTok{lslm}\NormalTok{( }\AttributeTok{y =}\NormalTok{ y, }\AttributeTok{x =}\NormalTok{ x, }\AttributeTok{weig =}\NormalTok{ weig, }\AttributeTok{boot =} \ConstantTok{TRUE}\NormalTok{ )}
\end{Highlighting}
\end{Shaded}

\begin{verbatim}
## [1] "------- Complete:20/200 -------"
## [1] "------- Complete:40/200 -------"
## [1] "------- Complete:60/200 -------"
## [1] "------- Complete:80/200 -------"
## [1] "------- Complete:100/200 -------"
## [1] "------- Complete:120/200 -------"
## [1] "------- Complete:140/200 -------"
## [1] "------- Complete:160/200 -------"
## [1] "------- Complete:180/200 -------"
## [1] "------- Complete:200/200 -------"
\end{verbatim}

If boot=TRUE, a nonparametric bootstrapping is performed to estimate the
95\% confidence intervals for the \(\tau^2\) and \(\rho\) parameters and
the direct and indirect effects, which quantify spill-over effects.
Default is FALSE. Here is the output in which \{s\_rho, sp\_SE\} are
parameters \{\(\rho, \tau^2\)\}:

\begin{Shaded}
\begin{Highlighting}[]
\NormalTok{res}
\end{Highlighting}
\end{Shaded}

\begin{verbatim}
## Call:
## lslm(y = y, x = x, weig = weig, boot = TRUE)
## 
## ----Coefficients------------------------------
##                  Estimate         SE    t_value      p_value
## (Intercept) -14.719039676 2.82212543 -5.2155866 2.748705e-07
## CRIM         -0.107615211 0.02851293 -3.7742599 1.809488e-04
## ZN            0.002594642 0.01276738  0.2032243 8.390474e-01
## INDUS        -0.098604511 0.06191541 -1.5925681 1.119273e-01
## CHAS          1.903178819 0.89128954  2.1353093 3.325050e-02
## NOX          -5.101316236 3.84673642 -1.3261414 1.854349e-01
## RM            6.922743307 0.33388005 20.7342228 0.000000e+00
## AGE          -0.040691404 0.01262483 -3.2231248 1.355874e-03
## 
## ----Spatial effects (lag)---------------------
##         Estimates    CI_lower   CI_upper
## sp_rho 0.02709059 -0.01858497 0.06728158
## sp_SD  7.54450065  6.41685806 8.65140381
## 
## ----Effects estimates-------------------------
## 
## Direct:
##          Estimates     CI_lower    CI_upper p_value
## CRIM  -0.107999852  -0.15893053 -0.04471181    0.00
## ZN     0.002603915  -0.01965664  0.02724361    0.74
## INDUS -0.098956945  -0.23073417  0.03676638    0.12
## CHAS   1.909981199   0.02430161  3.65799596    0.05
## NOX   -5.119549463 -13.81471930  2.31495559    0.20
## RM     6.947486715   6.28000232  7.68019902    0.00
## AGE   -0.040836844  -0.06540266 -0.01665769    0.00
## 
## Indirect:
##           Estimates     CI_lower     CI_upper p_value
## CRIM  -2.227815e-03 -0.005581694 0.0012902592    0.24
## ZN     5.371341e-05 -0.000620349 0.0006535612    0.86
## INDUS -2.041278e-03 -0.007718681 0.0018311012    0.34
## CHAS   3.939898e-02 -0.021583633 0.1275708059    0.29
## NOX   -1.056058e-01 -0.508362179 0.0794946205    0.38
## RM     1.433123e-01 -0.091783315 0.3587395220    0.24
## AGE   -8.423800e-04 -0.002284553 0.0005554187    0.24
## 
## ----Error statistics--------------------------
##                      stat
## resid_SE        3.9555161
## adjR2(cond)     0.8129243
## rlogLik     -1561.3219098
## AIC          3144.6438195
## BIC          3191.1357229
## 
## Note: The AIC and BIC values are based on the restricted likelihood.
##       Use method ="ml" for comparison of models with different fixed effects (x)
\end{verbatim}

\hypertarget{low-rank-spatial-error-model}{%
\subsubsection{Low rank spatial error
model}\label{low-rank-spatial-error-model}}

The low rank spatial error model (LSEM) approximates the following
model:
\[y_i=\beta_0+z_i+\epsilon_i\hspace{0.5cm}\epsilon_i \sim N(0,\sigma^2)\\
z_i=\sum^K_{k \neq 1}x_{i,k}\beta_k+e_i \hspace{0.5cm} e_i=\lambda\sum^N_{i \neq j}w_{i,j}e_j+ u_i \hspace{0.5cm} u_i \sim N(0,\tau^2)\]
where \(z_i\) is defined by the classical spatial error model (SLM) with
parameters \(\lambda\) and \(\tau^2\). Just like the original SEM,
\(\lambda\) takes a larger positive value in the presence of stronger
positive dependence. \(\tau^2\) represents the variance of the SEM-based
spatial process (i.e., \(z_i\)). As with LSLM, the LSEM estimates can be
different from the original SEM if data is noisy.

The lsem function estimates LSEM, as follows:

\begin{Shaded}
\begin{Highlighting}[]
\FunctionTok{data}\NormalTok{(boston)}
\NormalTok{res   }\OtherTok{\textless{}{-}} \FunctionTok{lsem}\NormalTok{( }\AttributeTok{y =}\NormalTok{ y, }\AttributeTok{x =}\NormalTok{ x, }\AttributeTok{weig =}\NormalTok{ weig )}
\NormalTok{res}
\end{Highlighting}
\end{Shaded}

\begin{verbatim}
## Call:
## lsem(y = y, x = x, weig = weig)
## 
## ----Coefficients------------------------------
##                  Estimate         SE    t_value      p_value
## (Intercept) -15.535928399 2.82054020 -5.5081393 6.082512e-08
## CRIM         -0.093112127 0.02911351 -3.1982447 1.479351e-03
## ZN            0.002300116 0.01292558  0.1779507 8.588411e-01
## INDUS        -0.063433279 0.06176206 -1.0270591 3.049394e-01
## CHAS          1.335521734 0.88216035  1.5139217 1.307414e-01
## NOX          -5.717186159 3.86329642 -1.4798725 1.396007e-01
## RM            7.052094665 0.33425292 21.0980796 0.000000e+00
## AGE          -0.037131943 0.01253448 -2.9623833 3.212894e-03
## 
## ----Spatial effects (residuals)---------------
##           Estimates
## sp_lambda  0.885701
## sp_SD      2.926975
## 
## ----Error statistics--------------------------
##                      stat
## resid_SE        4.0001174
## adjR2(cond)     0.8086816
## rlogLik     -1544.3307054
## AIC          3110.6614108
## BIC          3157.1533142
## 
## Note: The AIC and BIC values are based on the restricted likelihood. 
##       Use method ="ml" for comparison of models with different fixed effects (x)
\end{verbatim}

\{s\_lambda, sp\_SE\} are parameters \{\(\lambda, \tau^2\)\}.

\hypertarget{modeling-large-samples}{%
\section{Modeling large samples}\label{modeling-large-samples}}

\hypertarget{eigen-decomposition}{%
\subsection{Eigen-decomposition}\label{eigen-decomposition}}

The meigen function implements an eigen-decomposition that is slow for
large samples. For fast eigen-approximation, the meigen\_f function is
available. By default, this function approximates 200 eigenvectors; 200
is based on simulation results in Murakami and Griffith (2019a). The
computation is further accelerated by reducing the number of
eigenvectors. It is achieved by specifying enum by a number smaller than
200. While the meigen function took 243.8 seconds for 5,000 samples, the
meigen\_f took less than 1 second, as demonstrated below:

\begin{Shaded}
\begin{Highlighting}[]
\NormalTok{coords\_test     }\OtherTok{\textless{}{-}} \FunctionTok{cbind}\NormalTok{( }\FunctionTok{rnorm}\NormalTok{( }\DecValTok{5000}\NormalTok{ ), }\FunctionTok{rnorm}\NormalTok{( }\DecValTok{5000}\NormalTok{ ) )}
\FunctionTok{system.time}\NormalTok{( meig\_test200   }\OtherTok{\textless{}{-}} \FunctionTok{meigen\_f}\NormalTok{( }\AttributeTok{coords =}\NormalTok{ coords\_test ))[}\DecValTok{3}\NormalTok{]}
\end{Highlighting}
\end{Shaded}

\begin{verbatim}
## elapsed 
##    0.22
\end{verbatim}

\begin{Shaded}
\begin{Highlighting}[]
\FunctionTok{system.time}\NormalTok{( meig\_test100   }\OtherTok{\textless{}{-}} \FunctionTok{meigen\_f}\NormalTok{( }\AttributeTok{coords =}\NormalTok{ coords\_test, }\AttributeTok{enum=}\DecValTok{100}\NormalTok{ ))[}\DecValTok{3}\NormalTok{]}
\end{Highlighting}
\end{Shaded}

\begin{verbatim}
## elapsed 
##    0.06
\end{verbatim}

\begin{Shaded}
\begin{Highlighting}[]
\FunctionTok{system.time}\NormalTok{( meig\_test50    }\OtherTok{\textless{}{-}} \FunctionTok{meigen\_f}\NormalTok{( }\AttributeTok{coords =}\NormalTok{ coords\_test, }\AttributeTok{enum=}\DecValTok{50}\NormalTok{ ))[}\DecValTok{3}\NormalTok{]}
\end{Highlighting}
\end{Shaded}

\begin{verbatim}
## elapsed 
##    0.05
\end{verbatim}

On the other hand, the weigen function implements the ARPACK routine for
fast eigen-decomposition by default. The computational times with 5,000
samples and enum = 200 (default), 100, and 50 are as follows:

\begin{Shaded}
\begin{Highlighting}[]
\FunctionTok{system.time}\NormalTok{( weig\_test200   }\OtherTok{\textless{}{-}} \FunctionTok{weigen}\NormalTok{( coords\_test ))[}\DecValTok{3}\NormalTok{]}
\end{Highlighting}
\end{Shaded}

\begin{verbatim}
## elapsed 
##    7.09
\end{verbatim}

\begin{Shaded}
\begin{Highlighting}[]
\FunctionTok{system.time}\NormalTok{( weig\_test100   }\OtherTok{\textless{}{-}} \FunctionTok{weigen}\NormalTok{( coords\_test, }\AttributeTok{enum=}\DecValTok{100}\NormalTok{ ))[}\DecValTok{3}\NormalTok{]}
\end{Highlighting}
\end{Shaded}

\begin{verbatim}
## elapsed 
##   12.75
\end{verbatim}

\begin{Shaded}
\begin{Highlighting}[]
\FunctionTok{system.time}\NormalTok{( weig\_test50    }\OtherTok{\textless{}{-}} \FunctionTok{weigen}\NormalTok{( coords\_test, }\AttributeTok{enum=}\DecValTok{50}\NormalTok{ ))[}\DecValTok{3}\NormalTok{]}
\end{Highlighting}
\end{Shaded}

\begin{verbatim}
## elapsed 
##     0.7
\end{verbatim}

\hypertarget{parameter-estimation}{%
\subsection{Parameter estimation}\label{parameter-estimation}}

The basic ESF model is estimated computationally efficiently by
specifying fn = ``all'' in the esf function. This setting is acceptable
for large samples (Murakami and Griffith, 2019a). The resf and resf\_vc
functions estimate all the models explained above using a fast
estimation algorithm developed in Murakami and Griffith (2019b). They
are available for large samples (e.g., 100,000 samples). Although the
SF-UQR model requires a bootstrapping to estimate confidential intervals
for the coefficients, the computational cost for the iteration does not
depend on sample size. Therefore, it is available for large samples too.

\hypertarget{sub-model-aggregation-for-improved-scalability-in-terms-of-accuracy}{%
\subsection{Sub-model aggregation for improved scalability in terms of
accuracy}\label{sub-model-aggregation-for-improved-scalability-in-terms-of-accuracy}}

The spatial regressions implemented in this package rely on a low rank
approximation (i.e., approximation that considers only a limited number
of eigen-pairs). For large samples (e.g., n \textgreater{} 5,000), this
approximation can lead to an degeneracy/over-smoothing of SVCs that
decreases modeling accuracy. To address this problem, the
addlearn\_local function additionally learns local patterns in the SVCs
by aggregating/averating a model pre-estimated by the resf\_vc or
besf\_vc function with local SVC models, which are defined by
k-means-based spatial clusters each of which contains roughly 600
samples (see Murakami et al., 2023). The last line below is a sample
example for the additional learning:

\begin{Shaded}
\begin{Highlighting}[]
\FunctionTok{data}\NormalTok{(house)}
\NormalTok{dat0    }\OtherTok{\textless{}{-}} \FunctionTok{st\_as\_sf}\NormalTok{(house)}
\NormalTok{dat0    }\OtherTok{\textless{}{-}}\NormalTok{ dat0[dat0}\SpecialCharTok{$}\NormalTok{yrbuilt}\SpecialCharTok{\textgreater{}}\DecValTok{1950}\NormalTok{,]}
\NormalTok{dat     }\OtherTok{\textless{}{-}} \FunctionTok{data.frame}\NormalTok{(}\FunctionTok{st\_coordinates}\NormalTok{(dat0),dat0[,}\FunctionTok{c}\NormalTok{(}\StringTok{"price"}\NormalTok{,}\StringTok{"age"}\NormalTok{,}\StringTok{"rooms"}\NormalTok{,}\StringTok{"beds"}\NormalTok{,}\StringTok{"syear"}\NormalTok{)])}
\NormalTok{coords  }\OtherTok{\textless{}{-}}\NormalTok{ dat[ ,}\FunctionTok{c}\NormalTok{(}\StringTok{"X"}\NormalTok{,}\StringTok{"Y"}\NormalTok{)];}\FunctionTok{names}\NormalTok{(coords)}\OtherTok{\textless{}{-}}\FunctionTok{c}\NormalTok{(}\StringTok{"px"}\NormalTok{,}\StringTok{"py"}\NormalTok{)}
\NormalTok{y         }\OtherTok{\textless{}{-}} \FunctionTok{log}\NormalTok{(dat[,}\StringTok{"price"}\NormalTok{])}
\NormalTok{x       }\OtherTok{\textless{}{-}}\NormalTok{ dat[,}\FunctionTok{c}\NormalTok{(}\StringTok{"age"}\NormalTok{,}\StringTok{"rooms"}\NormalTok{,}\StringTok{"beds"}\NormalTok{,}\StringTok{"syear"}\NormalTok{)]}
\NormalTok{meig      }\OtherTok{\textless{}{-}} \FunctionTok{meigen\_f}\NormalTok{(}\AttributeTok{coords=}\NormalTok{coords)}
\NormalTok{res     }\OtherTok{\textless{}{-}} \FunctionTok{resf\_vc}\NormalTok{(}\AttributeTok{y=}\NormalTok{y,}\AttributeTok{x=}\NormalTok{x,}\AttributeTok{meig=}\NormalTok{meig )}
\end{Highlighting}
\end{Shaded}

\begin{verbatim}
## [1] "-------  Iteration 1  -------"
## [1] "1/5"
## [1] "2/5"
## [1] "3/5"
## [1] "4/5"
## [1] "5/5"
## [1] "BIC: 6934.685"
## [1] "-------  Iteration 2  -------"
## [1] "1/5"
## [1] "2/5"
## [1] "3/5"
## [1] "4/5"
## [1] "5/5"
## [1] "BIC: 6832.381"
## [1] "-------  Iteration 3  -------"
## [1] "1/5"
## [1] "2/5"
## [1] "3/5"
## [1] "4/5"
## [1] "5/5"
## [1] "BIC: 6816.822"
## [1] "-------  Iteration 4  -------"
## [1] "1/5"
## [1] "2/5"
## [1] "3/5"
## [1] "4/5"
## [1] "5/5"
## [1] "BIC: 6815.569"
## [1] "-------  Iteration 5  -------"
## [1] "1/5"
## [1] "2/5"
## [1] "3/5"
## [1] "4/5"
## [1] "5/5"
## [1] "BIC: 6815.497"
## [1] "-------  Iteration 6  -------"
## [1] "1/5"
## [1] "2/5"
## [1] "3/5"
## [1] "4/5"
## [1] "5/5"
## [1] "BIC: 6815.494"
## [1] "-------  Iteration 7  -------"
## [1] "1/5"
## [1] "2/5"
## [1] "3/5"
## [1] "4/5"
## [1] "5/5"
## [1] "BIC: 6815.494"
## [1] "-------  Iteration 8  -------"
## [1] "1/5"
## [1] "2/5"
## [1] "3/5"
## [1] "4/5"
## [1] "5/5"
## [1] "BIC: 6815.494"
\end{verbatim}

\begin{Shaded}
\begin{Highlighting}[]
\NormalTok{res2    }\OtherTok{\textless{}{-}} \FunctionTok{addlearn\_local}\NormalTok{(res)}
\end{Highlighting}
\end{Shaded}

\begin{verbatim}
## [1] "-------- Aggregating 21 local sub-models ---------"
## [1] "1/21"
## [1] "2/21"
## [1] "3/21"
## [1] "4/21"
## [1] "5/21"
## [1] "6/21"
## [1] "7/21"
## [1] "8/21"
## [1] "9/21"
## [1] "10/21"
## [1] "11/21"
## [1] "12/21"
## [1] "13/21"
## [1] "14/21"
## [1] "15/21"
## [1] "16/21"
## [1] "17/21"
## [1] "18/21"
## [1] "19/21"
## [1] "20/21"
## [1] "21/21"
\end{verbatim}

\begin{Shaded}
\begin{Highlighting}[]
\NormalTok{res2}
\end{Highlighting}
\end{Shaded}

\begin{verbatim}
## Call:
## addlearn_local(mod = res)
## 
## ----Spatially varying coefficients on x (summary)----
## 
## Coefficient estimates:
##   (Intercept)          age              rooms               beds        
##  Min.   :-87.59   Min.   :-2.7128   Min.   :-0.06609   Min.   :0.01934  
##  1st Qu.:-67.69   1st Qu.:-0.8465   1st Qu.: 0.07580   1st Qu.:0.01934  
##  Median :-64.21   Median :-0.5697   Median : 0.08623   Median :0.01934  
##  Mean   :-64.15   Mean   :-0.3953   Mean   : 0.08973   Mean   :0.01934  
##  3rd Qu.:-58.18   3rd Qu.:-0.2364   3rd Qu.: 0.09879   3rd Qu.:0.01934  
##  Max.   :-32.97   Max.   :17.3201   Max.   : 0.25136   Max.   :0.01934  
##      syear        
##  Min.   :0.03764  
##  1st Qu.:0.03764  
##  Median :0.03764  
##  Mean   :0.03764  
##  3rd Qu.:0.03764  
##  Max.   :0.03764  
## 
## Statistical significance:
##                         Intercept  age rooms  beds syear
## Not significant                 0 3046    50     0     0
## Significant (10% level)         0  447    60     0     0
## Significant ( 5% level)         0  952   151 12299     0
## Significant ( 1% level)     12299 7854 12038     0 12299
## 
## ----Variance parameters----------------------------------
## 
## Spatial effects (Local sub-models; Average):
##                      (Intercept)      age      rooms beds syear
## random_SD              0.1949318 1.114462 0.01614957    0     0
## Moran.I/max(Moran.I)   0.5438285 0.288286 0.20392920   NA    NA
## 
## Spatial effects (Global sub-model):
##                      (Intercept)        age      rooms beds syear
## random_SD             0.08269065 0.15023812 0.00737726    0     0
## Moran.I/max(Moran.I)  0.13719812 0.06259446 0.11833633   NA    NA
## 
## ----Error statistics-------------------------------------
##                      stat
## resid_SE        0.2651236
## adjR2(cond)     0.7611387
## rlogLik      1722.2272705
## AIC         -3004.4545410
## BIC         -1372.6544288
## 
## NULL model: lm( y ~ x )
##    (r)loglik: -6266.884 ( AIC: 12545.77,  BIC: 12590.27 )
\end{verbatim}

The addlearn\_local function can be paralleled by specifying parallel=
TRUE. The smaller error of res2 over res confirms its better accuracy:

\begin{Shaded}
\begin{Highlighting}[]
\NormalTok{res}\SpecialCharTok{$}\NormalTok{e }\CommentTok{\# Before the adjustment}
\end{Highlighting}
\end{Shaded}

\begin{verbatim}
##                      stat
## resid_SE        0.2992962
## adjR2(cond)     0.7007731
## rlogLik     -3351.2431146
## AIC          6726.4862292
## BIC          6815.4935080
\end{verbatim}

The plot\_s function is available to quickly visualize the estimated
SVCs:

\begin{Shaded}
\begin{Highlighting}[]
\FunctionTok{plot\_s}\NormalTok{(res2,}\DecValTok{2}\NormalTok{)}\DocumentationTok{\#\#\# coefficients on rooms}
\end{Highlighting}
\end{Shaded}

\includegraphics{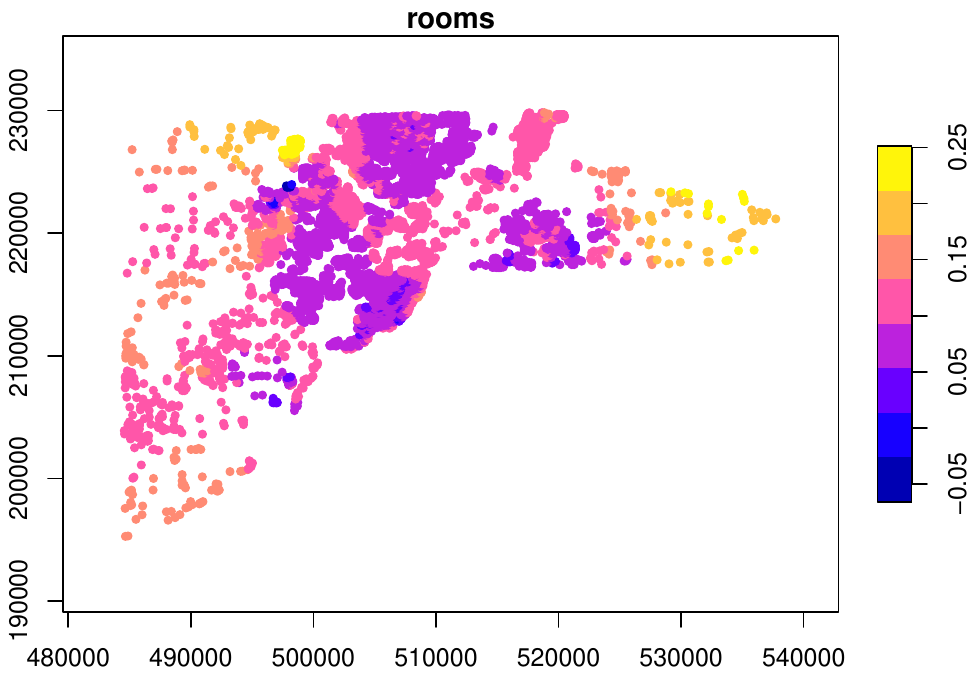}
The addlearn\_local function is also useful to improve predictive
accuracy for large samples. See Section 2.3 for further detail.

\hypertarget{for-very-large-samples-e.g.-n-100000}{%
\subsection{For very large samples (e.g., n \textgreater{}
100,000)}\label{for-very-large-samples-e.g.-n-100000}}

A computational limitation is the memory consumption of the meigen and
meigen\_f functions to store Moran eigenvectors. Because of the
limitation, the resf and resf\_vc functions are not available for very
large samples (e.g., millions of samples). To overcome this limitation,
the besf and besf\_vc functions perform the same calculation as resf and
resf\_vc but without saving the eigenvectors in the memory. Besides, for
fast computation, these functions perform a parallel model estimation
(see Murakami and Griffith, 2019c).

Here is an example implementing a spatial regression model using the
besf function and an SVC model using the besf\_vc function:

\begin{Shaded}
\begin{Highlighting}[]
\FunctionTok{data}\NormalTok{(house)}
\NormalTok{dat0  }\OtherTok{\textless{}{-}} \FunctionTok{st\_as\_sf}\NormalTok{(house)}
\NormalTok{dat0  }\OtherTok{\textless{}{-}}\NormalTok{ dat0[dat0}\SpecialCharTok{$}\NormalTok{yrbuilt}\SpecialCharTok{\textgreater{}}\DecValTok{1950}\NormalTok{,]}
\NormalTok{dat   }\OtherTok{\textless{}{-}} \FunctionTok{data.frame}\NormalTok{(}\FunctionTok{st\_coordinates}\NormalTok{(dat0),dat0[,}\FunctionTok{c}\NormalTok{(}\StringTok{"price"}\NormalTok{,}\StringTok{"age"}\NormalTok{,}\StringTok{"rooms"}\NormalTok{,}\StringTok{"beds"}\NormalTok{,}\StringTok{"syear"}\NormalTok{)])}
\NormalTok{coords}\OtherTok{\textless{}{-}}\NormalTok{ dat[ ,}\FunctionTok{c}\NormalTok{(}\StringTok{"X"}\NormalTok{,}\StringTok{"Y"}\NormalTok{)]}
\NormalTok{y       }\OtherTok{\textless{}{-}} \FunctionTok{log}\NormalTok{(dat[,}\StringTok{"price"}\NormalTok{])}
\NormalTok{x     }\OtherTok{\textless{}{-}}\NormalTok{ dat[,}\FunctionTok{c}\NormalTok{(}\StringTok{"age"}\NormalTok{,}\StringTok{"rooms"}\NormalTok{,}\StringTok{"beds"}\NormalTok{,}\StringTok{"syear"}\NormalTok{)]}
\NormalTok{res1    }\OtherTok{\textless{}{-}} \FunctionTok{besf}\NormalTok{(}\AttributeTok{y=}\NormalTok{y, }\AttributeTok{x=}\NormalTok{x, }\AttributeTok{coords=}\NormalTok{coords)}
\NormalTok{res1}
\end{Highlighting}
\end{Shaded}

\begin{verbatim}
## Call:
## besf(y = y, x = x, coords = coords)
## 
## ----Coefficients------------------------------
##                 Estimate          SE    t_value       p_value
## (Intercept) -59.64833595 3.486475791 -17.108490  1.282823e-65
## age          -0.49900873 0.032601324 -15.306394  6.930064e-53
## rooms         0.11115457 0.003845559  28.904661 1.043296e-183
## beds          0.01305038 0.006939198   1.880675  6.001615e-02
## syear         0.03532454 0.001746652  20.224145  6.002154e-91
## 
## ----Variance parameter------------------------
## 
## Spatial effects (residuals):
##                      (Intercept)
## random_SD             0.05277246
## Moran.I/max(Moran.I)  0.21609798
## 
## ----Error statistics--------------------------
##                      stat
## resid_SE        0.3139118
## adjR2(cond)     0.6709422
## rlogLik     -3624.8802859
## AIC          7265.7605719
## BIC          7325.0987578
## 
## Note: The AIC and BIC values are based on the restricted likelihood.
##       Use method ="ml" for comparison of models with different fixed effects (x)
\end{verbatim}

\begin{Shaded}
\begin{Highlighting}[]
\NormalTok{res2    }\OtherTok{\textless{}{-}} \FunctionTok{besf\_vc}\NormalTok{(}\AttributeTok{y=}\NormalTok{y, }\AttributeTok{x=}\NormalTok{x, }\AttributeTok{coords=}\NormalTok{coords)}
\end{Highlighting}
\end{Shaded}

\begin{verbatim}
## [1] "-------  Iteration 1  -------"
## [1] "1/5"
## [1] "2/5"
## [1] "3/5"
## [1] "4/5"
## [1] "5/5"
## [1] "BIC: 6603.706"
## [1] "-------  Iteration 2  -------"
## [1] "1/5"
## [1] "2/5"
## [1] "3/5"
## [1] "4/5"
## [1] "5/5"
## [1] "BIC: 6521.488"
## [1] "-------  Iteration 3  -------"
## [1] "1/5"
## [1] "2/5"
## [1] "3/5"
## [1] "4/5"
## [1] "5/5"
## [1] "BIC: 6518.712"
## [1] "-------  Iteration 4  -------"
## [1] "1/5"
## [1] "2/5"
## [1] "3/5"
## [1] "4/5"
## [1] "5/5"
## [1] "BIC: 6518.632"
## [1] "-------  Iteration 5  -------"
## [1] "1/5"
## [1] "2/5"
## [1] "3/5"
## [1] "4/5"
## [1] "5/5"
## [1] "BIC: 6518.631"
## [1] "-------  Iteration 6  -------"
## [1] "1/5"
## [1] "2/5"
## [1] "3/5"
## [1] "4/5"
## [1] "5/5"
## [1] "BIC: 6518.631"
\end{verbatim}

\begin{Shaded}
\begin{Highlighting}[]
\NormalTok{res2}
\end{Highlighting}
\end{Shaded}

\begin{verbatim}
## Call:
## besf_vc(y = y, x = x, coords = coords)
## 
## ----Spatially varying coefficients on x (summary)----
## 
## Coefficient estimates:
##   (Intercept)          age              rooms               beds        
##  Min.   :-61.45   Min.   :-3.2777   Min.   :-0.04103   Min.   :0.01528  
##  1st Qu.:-61.45   1st Qu.:-0.9951   1st Qu.: 0.08365   1st Qu.:0.01528  
##  Median :-61.45   Median :-0.6488   Median : 0.09439   Median :0.01528  
##  Mean   :-61.45   Mean   :-0.5265   Mean   : 0.09967   Mean   :0.01528  
##  3rd Qu.:-61.45   3rd Qu.:-0.1449   3rd Qu.: 0.10889   3rd Qu.:0.01528  
##  Max.   :-61.45   Max.   : 3.2467   Max.   : 0.25171   Max.   :0.01528  
##      syear        
##  Min.   :0.03552  
##  1st Qu.:0.03616  
##  Median :0.03630  
##  Mean   :0.03627  
##  3rd Qu.:0.03640  
##  Max.   :0.03656  
## 
## Statistical significance:
##                         Intercept  age rooms  beds syear
## Not significant                 0 5039   129     0     0
## Significant (10% level)         0  972   111     0     0
## Significant ( 5% level)         0 1762   385 12299     0
## Significant ( 1% level)     12299 4526 11674     0 12299
## 
## ----Variance parameters----------------------------------
## 
## Spatial effects (coefficients on x):
##                       (Intercept)        age       rooms beds        syear
## random_SD            5.546844e-05 0.17344865 0.007346195    0 3.496183e-05
## Moran.I/max(Moran.I) 9.146344e-01 0.06764239 0.107462739   NA 1.808831e-01
## 
## ----Error statistics-------------------------------------
##                      stat
## resid_SE        0.2930921
## adjR2(cond)     0.7130031
## rlogLik     -3193.3944912
## AIC          6414.7889823
## BIC          6518.6308077
## 
## Note: AIC and BIC are based on the restricted/marginal likelihood.
##       Use method="ml" for comparison of models with different fixed effects (x and xconst)
\end{verbatim}

Roughly speaking, these functions are faster than the resf and resf\_vc
functions if the sample size is more than 100,000.

As with the resf\_vc function, the besf\_vc function can suffer from the
degeneracy/over-smoothing problem. The addlearn\_local function is
useful to address this problem and improves SVC modeling accuracy:

\begin{Shaded}
\begin{Highlighting}[]
\NormalTok{res2b }\OtherTok{\textless{}{-}} \FunctionTok{addlearn\_local}\NormalTok{(res2)}
\end{Highlighting}
\end{Shaded}

\begin{verbatim}
## [1] "-------- Aggregating 23 local sub-models ---------"
\end{verbatim}

\begin{Shaded}
\begin{Highlighting}[]
\NormalTok{res2b}
\end{Highlighting}
\end{Shaded}

\begin{verbatim}
## Call:
## addlearn_local(mod = res2)
## 
## ----Spatially varying coefficients on x (summary)----
## 
## Coefficient estimates:
##   (Intercept)          age              rooms               beds        
##  Min.   :-84.42   Min.   :-4.6251   Min.   :-0.10442   Min.   :0.02006  
##  1st Qu.:-67.65   1st Qu.:-0.8730   1st Qu.: 0.07328   1st Qu.:0.02006  
##  Median :-63.99   Median :-0.6207   Median : 0.08105   Median :0.02006  
##  Mean   :-63.65   Mean   :-0.3669   Mean   : 0.08570   Mean   :0.02006  
##  3rd Qu.:-58.40   3rd Qu.:-0.2364   3rd Qu.: 0.09056   3rd Qu.:0.02006  
##  Max.   :-36.80   Max.   :22.3668   Max.   : 0.28422   Max.   :0.02006  
##      syear        
##  Min.   :0.02376  
##  1st Qu.:0.03455  
##  Median :0.03760  
##  Mean   :0.03741  
##  3rd Qu.:0.03962  
##  Max.   :0.04788  
## 
## Statistical significance:
##                         Intercept  age rooms  beds syear
## Not significant                 0 3142   116     0     0
## Significant (10% level)         0  685    78     0     0
## Significant ( 5% level)         0 1406   111 12299     0
## Significant ( 1% level)     12299 7066 11994     0 12299
## 
## ----Variance parameters----------------------------------
## 
## Spatial effects (Local sub-models; Average):
##                      (Intercept)       age      rooms beds        syear
## random_SD              0.2203258 1.6257069 0.01268354    0 9.803077e-05
## Moran.I/max(Moran.I)   0.5431389 0.2349645 0.15395648   NA 2.774002e-01
## 
## Spatial effects (Global sub-model):
##                       (Intercept)        age       rooms beds        syear
## random_SD            5.546844e-05 0.17344865 0.007346195    0 3.496183e-05
## Moran.I/max(Moran.I) 9.146344e-01 0.06764239 0.107462739   NA 1.808831e-01
## 
## ----Error statistics-------------------------------------
##                      stat
## resid_SE        0.2545360
## adjR2(cond)     0.7792871
## rlogLik      1167.8206794
## AIC         -1835.6413588
## BIC            18.6769505
## 
## NULL model: lm( y ~ x )
##    (r)loglik: -6266.884 ( AIC: 12545.77,  BIC: 12590.27 )
\end{verbatim}

The function is paralleled by default when besf\_vc is assumed.

\hypertarget{reference}{%
\section{Reference}\label{reference}}

\begin{itemize}
\tightlist
\item
  Chan, A. B., and Vasconcelos, N. (2011) Counting people with low-level
  features and Bayesian regression. IEEE Transactions on Image
  Processing, 21(4), 2160-2177.
\item
  Croissant, Y., and Millo, G. (2008) Panel data econometrics in R: The
  plm package. Journal of statistical software, 27(2), 1-43.
\item
  Firpo, S., Fortin, N.M., and Lemieux, T. (2009) Unconditional quantile
  regressions. Econometrica, 77 (3), 953-973.
\item
  Griffith, D.A. (2003) Spatial autocorrelation and spatial filtering:
  gaining understanding through theory and scientific visualization.
  Springer Science \& Business Media.
\item
  Ghosh, M., and Rao, J. N.K. (1994) Small area estimation: an
  appraisal. Statistical science, 9 (1), 55-76.
\item
  LeSage, J.P. and Pace, R.K. (2009) Introduction to Spatial
  Econometrics. CRC Press.
\item
  Murakami, D. and Griffith, D.A. (2015) Random effects specifications
  in eigenvector spatial filtering: a simulation study. Journal of
  Geographical Systems, 17 (4), 311-331.
\item
  Murakami, D. and Griffith, D.A. (2019a) Eigenvector spatial filtering
  for large data sets: fixed and random effects approaches. Geographical
  Analysis, 51 (1), 23-49.
\item
  Murakami, D. and Griffith, D.A. (2019b) Spatially varying coefficient
  modeling for large datasets: Eliminating N from spatial regressions.
  Spatial Statistics, 30, 39-64.
\item
  Murakami, D. and Griffith, D.A. (2019c) A memory-free spatial additive
  mixed modeling for big spatial data. Japan Journal of Statistics and
  Data Science, doi: 10.1007/s42081-019-00063-x.
\item
  Murakami, D., Griffith, D.A. (2021) Balancing spatial and
  non-spatially variations in varying coefficient modeling: a remedy for
  spurious correlation. Geographical Analysis,
  \url{DOI:10.1111/gean.12310}.
\item
  Murakami, D. (2021) Transformation-based generalized spatial
  regression using the spmoran package: Case study examples, ArXiv.
\item
  Murakami, D., Kajita, M., Kajita, S., Matsui, T. (2021)
  Compositionally warped additive modeling for a wide variety of
  non-Gaussian spatial data. Arxiv.
\item
  Murakami, D. and Seya, H. (2019) Spatially filtered unconditional
  quantile regression. Environmetrics, 30 (5), e2556.
\item
  Murakami, D., Seya, H., and Griffith, D.A. (2018) Low rank spatial
  econometric models. Arxiv, 1810.02956.
\item
  Murakami, D., Sugasawa, S., T., Seya, H., and Griffith, D.A. (2023)
  Sub-model aggregation-based scalable eigenvector spatial filtering:
  application to spatially varying coefficient modeling. Arxiv.
\item
  Murakami, D., Yoshida, T., Seya, H., Griffith, D.A., and Yamagata, Y.
  (2017) A Moran coefficient-based mixed effects approach to investigate
  spatially varying relationships. Spatial Statistics, 19, 68-89.
\item
  Rios, G., Tobar, F. (2019). Compositionally-warped Gaussian processes.
  Neural Networks, 118, 235-246.
\item
  Wheeler, D., and Tiefelsdorf, M. (2005) Multicollinearity and
  correlation among local regression coefficients in geographically
  weighted regression. Journal of Geographical Systems, 7(2), 161-187.
\item
  Yu, D., Murakami, D., Zhang, Y., Wu, X., Li, D., Wang, X., and Li, G.
  (2020) Investigating high-speed rail construction's support to county
  level regional development in China: An eigenvector based spatial
  filtering panel data analysis. Transportation Research Part B:
  Methodological, 133, 21-37.
\end{itemize}

\end{document}